\documentclass[amsmath,amssymb,twocolumn,aps,pre,twocolumn,superscriptaddress,english]{revtex4-1}  

\pdfoutput=1

\usepackage{lmodern}
\usepackage[T1]{fontenc}
\usepackage[latin9]{inputenc}
\usepackage{amsmath}
\usepackage{amssymb}
\usepackage{mathrsfs}
\usepackage{bm}
\usepackage{esint}
\usepackage{siunitx} 
\usepackage{xcolor}
\usepackage{babel}
\usepackage{float}
\usepackage{datetime}
\usepackage{dcolumn}   
\usepackage{graphicx}  
\usepackage{placeins}
\usepackage{hyperref}
\usepackage{float}

\hypersetup{colorlinks,bookmarksopen,bookmarksnumbered,
citecolor=blue,
linkcolor=black,
pdfstartview=green,
urlcolor=purple}

\newcommand{\D}{{\rm d}}

\newcommand{\qk}{Q_k}
\newcommand{\qkb}{Q_{\bar{k}}}

\newcommand{\lk}{L_k}
\newcommand{\lkb}{L_{\bar{k}}}

\def\be{\begin{equation}}
\def\ee{\end{equation}}
\def\bea{\begin{eqnarray}}
\def\eea{\end{eqnarray}}

\begin{document}

\title{Thermodynamic equilibrium of binary mixtures on curved surfaces}

\date{\today}

\author{Piermarco Fonda}
\email{fonda@lorentz.leidenuniv.nl}
\affiliation{Instituut-Lorentz, Universiteit Leiden, P.O. Box 9506, 2300 RA Leiden, The Netherlands}
\author{Melissa Rinaldin}
\affiliation{Instituut-Lorentz, Universiteit Leiden, P.O. Box 9506, 2300 RA Leiden, The Netherlands}
\affiliation{Huygens-Kamerlingh Onnes Lab, Universiteit Leiden, P. O. Box 9504, 2300 RA Leiden, The Netherlands}
\author{Daniela J. Kraft}
\affiliation{Huygens-Kamerlingh Onnes Lab, Universiteit Leiden, P. O. Box 9504, 2300 RA Leiden, The Netherlands}
\author{Luca Giomi}
\email{giomi@lorentz.leidenuniv.nl}
\affiliation{Instituut-Lorentz, Universiteit Leiden, P.O. Box 9506, 2300 RA Leiden, The Netherlands}

\begin{abstract}
We study the global influence of curvature on the free energy landscape of two-dimensional binary mixtures confined on closed surfaces. Starting from a generic effective free energy, constructed on the basis of symmetry considerations and conservation laws, we identify several model-independent phenomena, such as a curvature-dependent line tension and local shifts in the binodal concentrations. To shed light on the origin of the phenomenological parameters appearing in the effective free energy, we further construct a lattice-gas model of binary mixtures on non-trivial substrates, based on the curved-space generalization of the two-dimensional Ising model. This allows us to decompose the interaction between the local concentration of the mixture and the substrate curvature into four distinct contributions, as a result of which the phase diagram splits into critical sub-diagrams. The resulting free energy landscape can admit, as stable equilibria, strongly inhomogeneous mixed phases, which we refer to as ``antimixed'' states below the critical temperature. We corroborate our semi-analytical findings with phase-field numerical simulations on realistic curved lattices. Despite this work being primarily motivated by recent experimental observations of multi-component lipid vesicles supported by colloidal scaffolds, our results are applicable to any binary mixture confined on closed surface of arbitrary geometry. 
\end{abstract}

\maketitle

\section{Introduction}

Two-dimensional fluids represent a special class of materials, whose mechanical and thermodynamical properties are simultaneously simple and exotic. Their dynamics and thermodynamics can be considerably less involved compared to three-dimensional counterparts (see e.g. Ref. \cite{batchelor1967introduction}). Yet, being lower dimensional systems embedded in higher dimensional space, their geometry and topology may be non-trivial. This gives rise to a variety of phenomena where the static and dynamical configurations of the fluid conspire with the shape of the underlying substrate, resulting in a wealth of complex mechanical and thermodynamical behaviours, ranging from the proliferation of defects in two-dimensional liquid crystals and superfluids \cite{Bowick2009,Turner2010} to the emergence of topologically protected oceanic waves \cite{Delplace2017}.

Lipid membranes represent one of the most relevant and largely studied realizations of two-dimensional fluids confined on curved surfaces. Artificial lipid membranes, i.e. \textit{in vitro} bilayers which have been purified from other components, have served for decades as fruitful model systems to investigate the stability and material properties of self-assembled biological lipid structures (see e.g. \cite{lipowsky1995structure,david2004statistical}). This is especially true in the case of artificial bilayers consisting of multiple lipid components (see e.g. \cite{Feigenson2009} and references therein), where the heterogeneity of the system shortens the gap between artificial and cellular membranes, despite maintaining a physically tractable complexity.

It is well-known that multi-component mixtures of phospholipids and cholesterol have rich phase diagrams, including two different types of liquids known as the (cholesterol-rich) liquid-ordered (LO) and liquid-disordered (LD) phases. While binary lipid mixtures, which provide the simplest example of a multi-component membrane, clearly exhibit coexistence between liquid and solid phases  \cite{mouritsen2015life}, there is still lack of conclusive evidence in support of a genuine LO/LD coexistence in mixtures of saturated lipids and cholesterol \cite{Marsh2010}. For this reason, and because liquid/liquid phase separation is believed to be very relevant for biological systems \cite{Stillwell2013}, most literature shifted the attention toward ternary membranes, usually featuring saturated and unsaturated lipids and cholesterol, where the critical nature of the phase separation is unquestioned. The LO/LD coexistence has so far been realized in several experimental set-ups which have also shown a a correlation between geometry and chemical composition: giant unilamellar vesicles (GUVs) \cite{Baumgart2003,Hess2007, Semrau2009, Sorre2009, Heinrich2010}, supported lipid bilayers (SLBs) \cite{Parthasarathy2006,Subramaniam2010} and scaffolded lipid vesicles (SLVs) \cite{rinaldin2018geometric}. 

Here we focus on SLVs, since it is the only experimental set-up that is simultaneously closed (i.e. as in GUVs, there is no exchange of lipids with the surrounding solvent) and of prescribed shape (a property shared with SLBs). We stress, nonetheless, that the results of the present work apply, in principle, to any generic two-dimensional liquid mixture confined on a curved substrate. SLVs have typical size of a few micrometers \cite{rinaldin2018geometric}, whereas a single lipid molecule occupies an area on the membrane of order $\sim 1$ \si{\nano\squared\metre} \cite{mouritsen2015life}: therefore, the number of constituents is approximately in the millions. With such a high number of molecules, it is natural to describe the membrane as a single smooth surface where the local composition is a continuous space-dependent field. A satisfactory physical description can be attained by a coarse-grained two-dimensional scalar field theory, with the fields representing the concentration of the various molecule types. For incompressible liquids, an $n$-component mixture is described by $n-1$ fields. 

Much of this work will focus on the ability of a single scalar field, $\phi$, to describe the curvature-composition interactions.
Despite being appropriate for binary systems only, a single scalar degree of freedom can capture, at least qualitatively, the effect of geometry on the structure of the free energy landscape and the resulting phase behaviour. Furthermore, focusing on a single field has numerous advantages, as it roots in the classical theory of phase separations and was first used to model the interaction between curvature and lipid lateral organization by Markin \cite{Markin1981} and Leibler \cite{Leibler1986}. In the latter work, the interplay between the membrane chemical composition and geometry was modelled in terms as a concentration-dependent spontaneous mean curvature, leading to a linear coupling in the effective free energy, analogous to that between an order parameter and an external ordering field. Such coupling breaks the reflection symmetry along the membrane mid-surface, since the mean curvature is sensitive to the surface orientation.

This type of interaction was adopted by many subsequent works (see e.g. Refs. \cite{Leibler1987,Taniguchi1996,Jiang2000,Gozdz2006,Singh2012,Zhu2012,schick2012membrane,Rautu2015}), whereas others (e.g. Refs. \cite{McWhirter2004,Baumgart2011,Gozdz2012}) considered also linear couplings with the squared mean curvature, which is better suited to describe symmetric bilayers. Conversely, other works did not introduce any interaction terms, but rather studied the effects of a non-trivial intrinsic geometry \cite{novick1991stable,Rubinstein1992,Gomez2015}. Explicit intrinsic couplings were considered  in Ref. \cite{paillusson2016phase}, with a direct coupling to the Gaussian curvature, and in Ref. \cite{adkins2017geodesic}, where the notion of spontaneous geodesic curvature was introduced. Note that, because of the Gauss-Bonnet theorem, a direct coupling between the Gaussian curvature and the concentration is irrelevant for chemically homogeneous membranes, and likely for this reason it has often been disregarded. Couplings quadratic and cubic in $\phi$ were considered in other works (e.g. in Refs. \cite{Jiang2008,Wang2008,Lowengrub2009,Elliott2010,Jiang2012,Choksi2013,Helmers2013,hausser2013thermodynamically}, and also by us in Ref. \cite{rinaldin2018geometric}) and appears to be the most popular choice within the mathematics-oriented literature. 

There is no general consensus on how to choose neither the type nor the functional form of the couplings between the shape and the concentration. Although linear terms are the natural choice from a field-theoretic point of view, it is not clear how model-specific will be the results obtained, and thus it is hard do assess their general validity. Furthermore, most of the cited works focus on the local and dynamical effect of given couplings in an open setting. However, vesicle-shaped objects are inherently constrained systems, being topologically spherical and with no relevant exchange with the surrounding environment: the total number of molecules is an externally fixed parameter. For these reasons, we try to have a more systematic approach and explore the all the possible equilibrium configurations of closed two-dimensional systems. For sake of conciseness, we ignore the role of fluctuations (but see e.g. Ref. \cite{david2004statistical}).

The paper is organized as follows.
In Sec. \ref{sec:effective} we develop an effective scalar field theory on curved backgrounds, using only symmetry and scaling arguments as guiding principles. We highlight a few possible general phenomena, such as local shifts of the binodal concentrations and a curvature-dependent line tension for interfaces separating different phases, and highlight the regimes where J\"ulicher's and Lipowsky's sharp interface theory \cite{Julicher1993} can be recovered from our diffuse interface model. In Sec. \ref{sec:model} we explore in great detail a specific geometry, the asymmetric dumbbell, and a specific microscopic model, consisting of a curved-space generalization of the mean-field two-dimensional Ising model. In the continuum limit, we derive a functional form of the concentration-dependent coefficients of the free energy, linking them to four specific types of microscopic interactions. Within this framework we can compute analytically the general quantities defined in Sec. \ref{sec:effective}. By approximating the dumbbell with two disjoint spheres able to exchange molecules, we construct temperature-concentration phase diagrams for any value of the curvature couplings. Interestingly, we are able to give a precise, model-independent definition of the antimixed state, which we observed experimentally in Ref. \cite{rinaldin2018geometric}. Lastly, we prove numerically that our results, and in particular the existence of the antimixed state, are robust and continue to apply also to more realistic geometries. 

\section{Mixing and demixing on curved surfaces}
\label{sec:effective}

\subsection{Effective free energies for inhomogeneous systems}
\label{sec:effective_intro}

We consider a two-dimensional binary fluid and assume that all the relevant degrees of freedom can be captured by a single, generally space-dependent, scalar order parameter $\phi=\phi(\bm{r})$. If the fluid is incompressible and the average area per molecule is the same for both components, $\phi$ can be interpreted as the absolute concentration of either one of the two components, e.g.:
\begin{equation}
\phi = \frac{[A]}{[A]+[B]}\;,
\end{equation} 
where $[\ldots]$ indicate the concentration of the $A$ and $B$ molecules. By construction, $0\le \phi \le 1$ and any value other than $\phi=0$ or $\phi=1$ indicates local mixing of the two components. The system is defined on an arbitrarily curved surface $\Sigma$. Crucially, we assume $\Sigma$ fixed so that the local geometry can influence the configuration of the order parameter $\phi$, but not vice versa. The most general free energy functional of such a system will then be of the form:
\be 
F = \int_\Sigma \D A\,\mathcal{F}(\phi,\nabla \phi, \Sigma) \;,
\label{effective F}
\ee
where $\mathcal{F}$ is a free energy density depending on $\phi$, its surface-covariant gradient $\nabla \phi$ and on the shape of the surface. Here $\D A = \D x^{1} \D x^{2} \sqrt{\det h}$, with $\{x^{1},x^{2}\}$ local coordinates, is the surface area element and $h_{ij}$ ($i,\,j=1,\,2$) is the metric tensor on $\Sigma$. 

In practice, the explicit form of $\mathcal{F}$ can be obtained upon coarse-graining a microscopic model over a mesoscopic portion of $\Sigma$. Such portion should be small compared to the size of the whole system and yet large compared to typical molecular length-scales, which we call $a$. Alternatively, as in most cases of practical interest, $\mathcal{F}$ is constructed phenomenologically, on the basis of symmetry arguments and physical insight. 
Because the order parameter is generally non-uniform across the surface, the gradient $\nabla \phi$ introduces new length scales in the system. Here we assume that the spatial variation of the order parameter occurs on a length scale much larger than the molecular size, namely: $|\nabla\phi| \sim \xi^{-1} \ll a^{-1}$. Moreover, at physical equilibrium, gradients are always negligible with the only possible exception for isolated quasi-one-dimensional regions where the spatial variation of the order parameter can be more pronounced. As we will explain later, these regions correspond to diffuse interfaces between bulk phases and, being lower dimensional structures, do not affect the bulk value of the free energy. Since integrated variations have to be finite, $\xi$ also sets the typical thickness of these interfaces.

The symmetries of Eq. \eqref{effective F} dictate how $\mathcal{F}$ can depend on the shape of $\Sigma$. If the fluid is isotropic (i.e. molecules do not have a specific direction on the tangent plane of $\Sigma$), $\mathcal{F}$ depends on the surface either intrinsically, through the Gaussian curvature $K$, or extrinsically, through the mean curvature $H$. Furthermore, if the molecules are insensitive to the orientation of the surface (i.e. they do not discriminate convex from concave shapes), $\mathcal{F}$ must be invariant for $H\rightarrow-H$, since, on orientable surfaces, the sign of $H$ depends uniquely on the choice of the normal direction. Thus $\mathcal{F}$ depends on the curvature only through $H^2$, $K$ and, in principle, their derivatives. Non-vanishing curvatures introduce further length scales in the system, which we collectively denote as $R$ and assume larger or equal to $\xi$, thus $R \geq \xi \gg a$.

Now, expanding Eq. \eqref{effective F} to the second order in the gradients and the curvatures (thus with respect to $a/\xi$ and $a/R$) yields:
\be
\mathcal{F} \simeq \frac{D(\phi)}{2}\,|\nabla\phi|^{2} +  f(\phi) + k(\phi) H^2 + \bar{k}(\phi) K  + \cdots \,,
\label{F expansion}
\ee
where $D$, $f$, $k$ and $\bar{k}$ are the resulting coefficients in the Taylor expansion and the dots indicate higher order terms. These coefficients depend, in general, on the local order parameter $\phi$ and cannot be determined from symmetry arguments. To render Eq. \eqref{F expansion} dimensionless, we rescale all the terms by a constant energy density, in such a way that $f$ is dimensionless, whereas $D$, $k$ and $\bar{k}$ have dimensions of area.

The physical meaning of the various terms in Eq. \eqref{F expansion} is intuitive and has been thoroughly discussed in the literature of phase field models \cite{Lowengrub2009,Elliott2010} and lipid membranes \cite{Lazaro2015}. To have an energy bounded from below requires $D \ge 0$, so that the first term promotes uniform configurations of the order parameter. This term originates from the short-range attractive interactions between molecules and gives rise to a concentration-dependent diffusion coefficient (see e.g. \cite{crank1979mathematics,Rubinstein1992,Schoenborn1997,Jiang2012}). 
Notice that $D$ does not depend on the curvatures, because of the quadratic truncation underling Eq. \eqref{F expansion}. 
Higher order terms coupling the order parameter gradients and the curvature tensor have been discussed elsewhere  (see e.g. \cite{Goetz1996,Fournier1997,Siegel2010,Deserno2015}) and will not be considered here. 
The function $f$ is the local thermodynamic free energy in flat space. This includes both energetic and entropic contributions, promoting phase separation and phase mixing respectively. 
In the case of fluctuating surfaces, such as lipid membranes, $f$ could be interpreted as a concentration-dependent surface tension. Finally, $k$ and $\bar{k}$ are, respectively, the bending and saddle splay moduli of the mixture, expressing the energetic cost, or gain, of having a given configuration of the field $\phi$, in a curved region of the surface. Analogously to $f$, for a fluctuating surface these terms could be interpreted as a curvature-dependent contributions to the surface tension, introducing a departure for the flat-space value. The length scale associated to these deviations is commonly known as Tolman length \cite{Tolman1949}. A generic surface may have up to two independent Tolman lengths.

For systems sensitive to the orientation of the surface, such as Langmuir monolayers and asymmetric lipid bilayers, the expansion \eqref{F expansion} is not required to be invariant for $H \rightarrow -H$ and can feature linear contributions of the form $c H$, with $c=c(\phi)$ a coupling coefficient, equivalent to a concentration-dependent spontaneous curvature $H_{0}=-c/(2k)$. For simplicity, we will ignore this contribution, even if most of our results can be easily extended to this case.

Equilibrium configurations are defined as the minima of the free energy functional Eq. \eqref{effective F}. Here we focus on closed systems, where the order parameter is globally conserved. Thus:
\be
\Phi = \frac{1}{A_{\Sigma}} \int_{\Sigma} \D A\,\phi = {\rm const}\,,
\label{Phi}
\ee
with $A_{\Sigma}$ the area of the surface. The problem then reduces to finding the function $\phi$ minimizing the constrained free energy:
\be
G = F - \hat{\mu} \Phi \;, 
\label{grand canonical G}
\ee
where $\hat{\mu}$ is the Lagrange multiplier enforcing the constraint \eqref{Phi}. For homogeneous systems, $\hat{\mu}$ is the chemical potential, thermodynamic conjugate of the concentration. The first functional derivative of $G$ yields the equilibrium condition
\begin{multline} 
f'(\phi) + k'(\phi) H^2 + \bar{k}'(\phi)K\\[5pt] = D(\phi) \nabla^{2} \phi + \frac{1}{2} D'(\phi) |\nabla\phi|^{2} + \mu \,,
\label{G variation}
\end{multline}
where the prime indicates differentiation with respect to $\phi$ (e.g. $f'=\partial f/\partial \phi$), $\nabla^{2}=h^{ij}\nabla_{i}\nabla_{j}$ is the Laplace-Beltrami operator on $\Sigma$ and $\mu=\hat{\mu}/A_\Sigma$ is the chemical potential density. Eq. \eqref{G variation} is too generic to draw specific conclusions, unless the $\phi-$dependence of the various coefficients is specified. In Sec. \ref{sec:model} we consider a specific lattice model, but, before then, it is useful to review the case of homogeneous potentials and make some general consideration on the linearization of inhomogeneous terms.

\subsection{Review of homogeneous potentials}
\label{sec:review}

\begin{figure}
\includegraphics[width=\columnwidth]{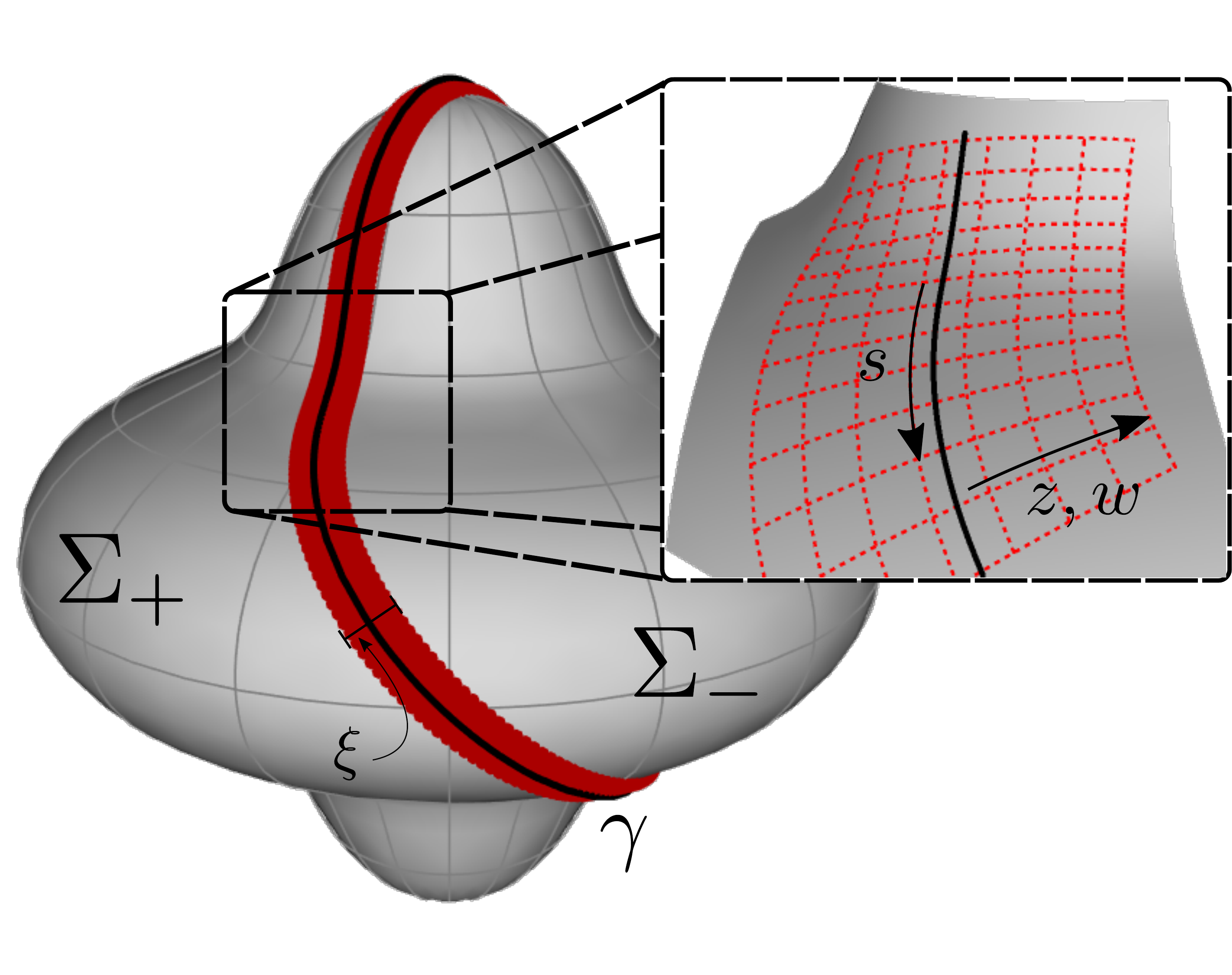}
\caption{When the system phase-separates on a SLV \cite{rinaldin2018geometric}, the surface $\Sigma$ (here shown as a generic closed surface) is partitioned into two regions $\Sigma_\pm$. The thin interface $\gamma$ separating them is a curved strip of finite geodesic width $\sim 2 \xi$, shown in red. In reality we require $\xi$ to be much smaller than any macroscopic length scale. In order to study the behaviour of $\phi(x)$ near the interface, we need to construct an adapted geodesic frame, spanned by the coordinates $s$, the arc-length parameter of the sharp interface (shown in black), and by the normal arc-length coordinate $z=w/\xi$. Constant $s$ lines are geodesics of $\Sigma$. }
\label{fig:figure 1}
\end{figure}

In this Section we review the classical theory of phase coexistence and of thin interfaces for binary mixtures in homogeneous backgrounds. For further references see e.g. Refs. \cite{Elder2001,provatas2011phase}.

\begin{figure}
\includegraphics[width=\columnwidth]{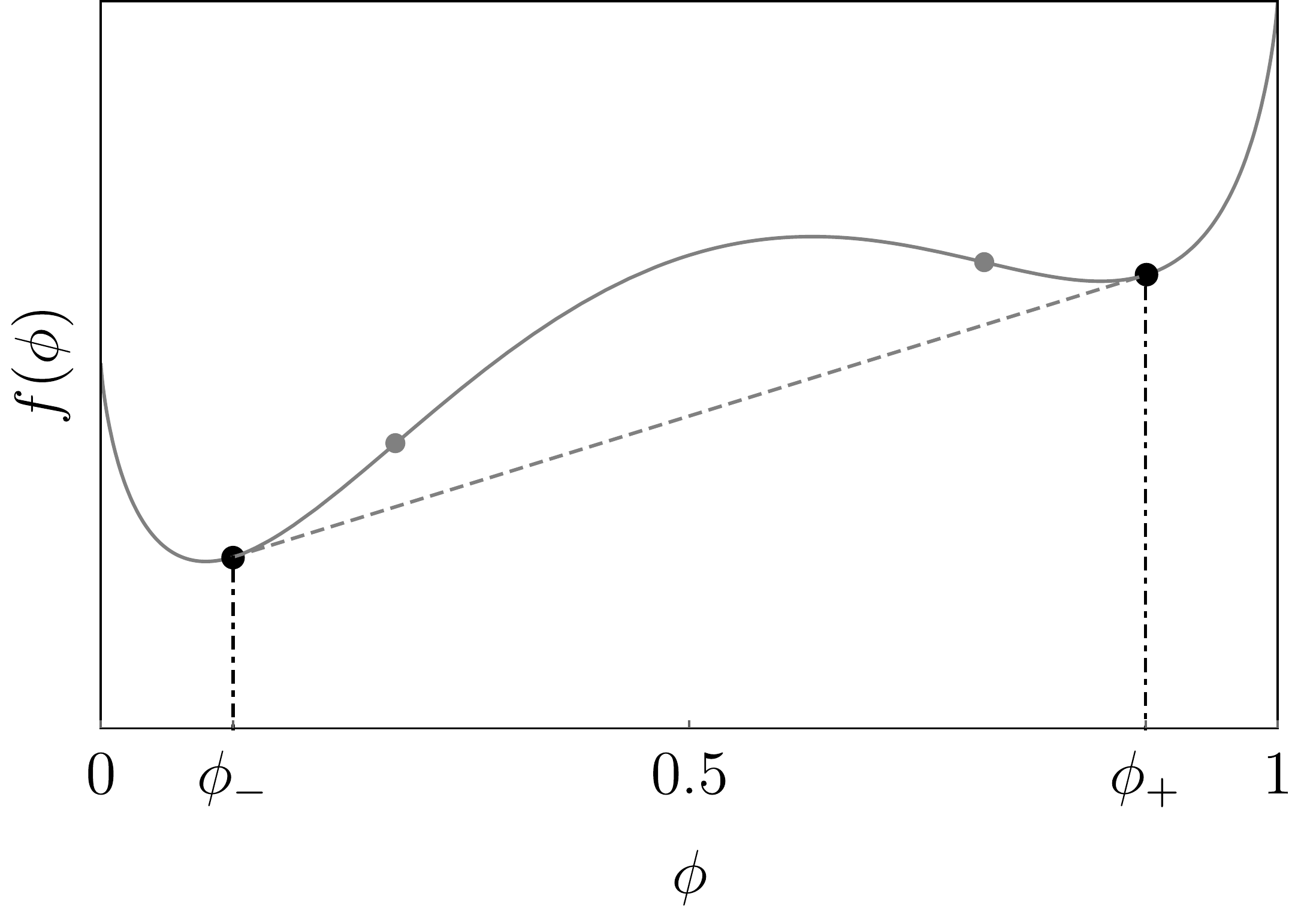}
\caption{For concave free energies the thermodynamic minimum is attained by demixed configurations when the total concentration $\Phi$ lies within the miscibility gap. We show respectively in black and gray the binodal and spinodal points relative to $f(\phi)$. The diagonal dashed line is the common tangent which defines, via Eq. \eqref{maxwell}, the binodal points. For a given $\Phi$, the area fractions $x_\pm$ of the $A$ and $B$ components are found with the lever rule, i.e. by solving \eqref{total phi} combined with $x_++x_-=1$.}
\label{fig:binodal}
\end{figure}

We now consider a flat and compact surface $\Sigma$, such as a rectangular domain with periodic boundaries (i.e. a flat torus). Thus $H^{2}=K=0$, while the total area $A_\Sigma$ is finite. Since $D \geq 0$, the homogeneous configuration is a trivial minimizer of the free energy \eqref{grand canonical G}. In most physical systems at equilibrium, field variations occur in almost-negligible portions of $\Sigma$, so that, as first crude approximation, gradient terms in $G$ can be ignored. Then, Eq. \eqref{G variation} reduces to the classical equilibrium condition
\be
f'(\phi) =\mu  \,.
\label{equilibrium 1}
\ee
If $f$ is convex, the single homogeneous phase $\phi=\Phi$ is a solution of Eq. \eqref{equilibrium 1}, corresponding to a stable thermodynamic state, where the two components of the mixture are homogeneously mixed with one another. We refer to this configuration as the mixed phase. Consistently we must have
\[
\mu=f'(\Phi)\;,\qquad \frac{G}{A_\Sigma} = f(\Phi)-f'(\Phi)\Phi\;.
\]
If, on the other hand, $f$ is concave for some $\phi$ values (i.e. $f''<0$), then the mixed phase might become unstable and it is energetically favourable to split the system into (at least) two regions where $\phi$ takes different values, say $\phi_-$ and $\phi_+$ (without loss of generality we choose $\phi_- < \phi_+$). We refer to this configuration as demixed (or phase-separated) phase:
\be 
\phi(\bm{r}) =
\left\{
\begin{array}{ll}
\phi_+,\;\; &\bm{r} \in \Sigma_+ \\
\phi_-,\;\; &\bm{r} \in \Sigma_-
\end{array}
\right.\,,
\label{sharp interface}
\ee
with $\Sigma_\pm$ the two domains into which $\Sigma$ partitions (see Fig. \ref{fig:figure 1}). Now, calling $A_\pm=\int_{\Sigma_{\pm}} \D A\;$ the respective areas and $x_\pm = A_\pm/ A_\Sigma$ their relative area fraction, with $x_{+}+x_{-}=1$, the total fixed concentration is
\be
\Phi = x_+ \phi_+ + x_- \phi_- \;.
\label{total phi}
\ee
Since $\phi$ is assumed to vary smoothly over a region of negligible area, it is possible to formally integrate Eq. \eqref{equilibrium 1} with respect to $\phi$ and obtain the set of  equilibrium conditions
\be
\mu = f'(\phi_\pm) = \frac{f(\phi_+)-f(\phi_-)}{\phi_+ - \phi_-} \;,
\label{maxwell}
\ee
known as Maxwell common-tangent construction, see Fig. \ref{fig:binodal}. The interval of $\Phi$ values for which the demixed phase, Eq. \eqref{sharp interface}, is the true minimum of the free energy \eqref{grand canonical G} is always strictly larger than the interval where $f(\Phi)$ is concave. Thus, a mixed phase with total concentration $\Phi$ in the interval $\phi_-< \Phi <\phi_+$, but such that $f''(\Phi)>0$, is metastable, since such phase can still resist small perturbations. The field values $\phi_\pm$ are known as binodal points, the interval $[\phi_-,\phi_+]$ is known as the miscibility gap, whereas the concentrations for which $f''(\phi)=0$ are known as spinodal points (see e.g. \citep{Safran1994}).

A further layer of complexity is added if one allows for smooth spatial variations of the order parameter $\phi$. In this case, the gradient terms in $G$ becomes relevant, but, because of the scale separation postulated in Sec. \ref{sec:effective_intro}, $D|\nabla \phi|^{2} \ll f$, almost everywhere. Since $|\nabla \phi| \sim \xi^{-1}$, by construction, and $D$ has dimensions of area in our units, this inequality implies $D \sim \xi^{2}$. We assume that $D$ - which relates to both compressibility and diffusion - does not depend strongly on the local concentration (for instance,  this is certainly the case for lipid mixtures \cite{Machan2010}, where all molecules in the mixture are roughly of the same size) and can be effectively treated as a constant. Without loss of generality, one can then set $D=\xi^{2}$, so that Eq. \eqref{equilibrium 1} reduces to partial differential equation:
\be
f'(\phi) = \mu + \xi^{2} \nabla^{2} \phi \;.
\label{equilibrium 2}
\ee
Since $f'$ is, in general, a non-linear function of $\phi$, Eq. \eqref{equilibrium 2} is often analytically intractable.
However, as long as $\xi$ is much smaller than the system size, Eq. \eqref{sharp interface}, is still a valid solution over large portions of $\Sigma$. Globally, the solution can then be constructed upon matching homogeneous configurations of the field over different domains of $\Sigma$ via perturbative solutions of Eq. \eqref{equilibrium 2} within the boundary layers at the interface between neighbouring domains (see e.g. Ref. \cite{provatas2011phase}). This is a standard technique which can be easily generalized to the case of curved environments, see also \cite{Rubinstein1992}.

\subsection{The effect of curvature}
\label{sec:curvature}

We now consider the more generic case in which $\Sigma$ has non-vanishing curvatures $H$ and $K$, but no explicit coupling with the order parameter (a similar situation in dynamical contexts was considered in Refs. \cite{Rubinstein1992,Gomez2015}), by setting $k=\bar{k}=0$ in Eq. \eqref{F expansion}. This scenario occurs, for instance, in mixtures whose components are equally compliant to bending, thus there is no energetic preference for the order parameter $\phi$ to adjust to the underlying curvature of the surface. Yet, as any interface in the configuration of the field $\phi$ costs a finite amount of energy, roughly proportional to the interface length, the shape of $\Sigma$ indirectly affects the spatial organization of the binary mixture via the geometry of interfaces. In Ref. \cite{fonda2018interface}, we have discussed this and other related phenomena in the framework of the sharp interface limit (i.e. with $\xi = 0$). Here we show how the present field-theoretical approach enables one to recover and further extend these results.

Upon demixing, the system drives the formation of interfaces. This means that in regions of thickness $\approx \xi$ the field $\phi$ is smoothly interpolating between the bulk values of the two phases. Since we are in a regime where this thickness is much smaller than the size of the system, we can take Eq. \eqref{equilibrium 2} and expand it in powers of $\xi$. As shown in Fig. \ref{fig:figure 1}, in the proximity of $\gamma$ we need to adapt the coordinate system to take into account both the curvature of the interface, as a strip embedded on the surface, and of the intrinsic curvature of the surface itself. We explain in detail how to build such frame in Appendix \ref{app:normal}. Then, we can treat the scalar field as a function of coordinates in this frame, $\phi=\phi(s,z)$, where $s$ is the arc-length parameter of the sharp interface $\gamma$ (the black curve in Fig. \ref{fig:figure 1}) and $z$ is the normal geodesic distance from the curve. Furthermore, variations along $z$ happen on a scale $\sim \xi$, while variations along $s$ become relevant only at macroscopic distances. This implies that $\phi$ is a function of only the normal coordinate $z$ up to at least order $\xi^2$, and we can rescale the variable $z \to w/\xi$ so that the values $w \to \pm \infty$ correspond to the bulk phases.

With this construction at hand, we collect the various terms in \eqref{equilibrium 2}, order by order in $\xi$, and solve iteratively the differential equation. At $O(1)$ we find the so-called profile equation, which, after matching with the bulk values of $\phi$ away from the interface, reads
\be
\frac{1}{2} \varphi_{w}^2 = g(\varphi)  \,,
\label{equipartition}
\ee
where $\varphi(w)=\phi(z/\xi)$ is the order parameter expressed as function of the rescaled normal coordinate $w$, and $g$ is the shifted potential
\be 
g(\varphi) = f(\varphi) + \frac{(\phi_- -\varphi) f(\phi_+)- (\phi_+-\varphi) f(\phi_-)}{\phi_+-\phi_-} \,,
\label{shifted g potential}
\ee
which has the properties $g(\phi_\pm)=g'(\phi_\pm)=0$ and $g''(\varphi)=f''(\varphi)$, i.e. $g$ shares the same binodal points with $f$.  Typically, solutions of Eq. \eqref{equipartition} decay exponentially towards the bulk phases and interpolate monotonically between the two phases. As we shall later see, this will not necessarily be the case for non-homogeneous systems.

Solving Eq. \eqref{equilibrium 2} at $O(\xi)$ is slightly more involved (see Appendix \ref{app:thin} for more details), but leads to a series of simple  and interesting results. First, in regions where $\xi^2 K$ is small, the equilibrium interface must obey
\be
\kappa_g = \rm{const}\,,
\label{CGC}
\ee
with $\kappa_g$ the geodesic curvature of the interface $\gamma$ (see Appendix \ref{app:normal} for definitions). Eq. \eqref{CGC} is the simplest two-dimensional version of the Young-Laplace equation on a curved geometry. The value of the constant, which is proportional to the lateral pressure difference on the two sides of the interface, sets the radius of curvature of the interface. While on a flat plane constant $\kappa_g$ lines are circles (and geodesics are straight lines), on an arbitrary surface they can have significantly less trivial shapes. We explored this subject in much more detail in \cite{fonda2018interface}, and refer the interested reader there. Note that \eqref{CGC} does not constrain the topology of $\gamma$: in principle it could consist of many simple curves, provided they all have the same curvature. In this case, the constraint on $\kappa_g$ is non-local \cite{Rubinstein1992}. 

From this it can be shown that a non geodesic interface induces a modification of the equilibrium chemical potential
\be
\mu = \frac{f(\phi_+)-f(\phi_-)}{\phi_+ - \phi_-} +  \frac{\sigma \kappa_g}{\phi_+-\phi_-} \,,
\label{sigma oxi}
\ee
where we introduced the interfacial line tension $\sigma$, defined as
\be 
\sigma = \xi \int_{\phi_-}^{\phi_+} \D \varphi  \sqrt{2 g(\varphi)} \,.
\label{line tension}
\ee
Eq. \eqref{sigma oxi} implies that, for non-geodesic interfaces, equilibrium bulk concentrations slightly deviate from the Maxwell values. 
This phenomenon is entirely absent in phase separations of open systems, where instead the bulk phases concentrations are not affected by the interface curvature. 

Such an effect is manifest also when evaluating the equilibrium free energy up to $O(\xi)$. Namely, we find 
\be
F = \sigma \ell_\gamma + \sum_{\alpha=\pm} A_\alpha \left[ f(\phi_\alpha) + \sigma \kappa_g \frac{f'(\phi_\alpha)}{(\phi_+-\phi_-)f''(\phi_\alpha)} \right] \,.
\label{thin F}
\ee
The above relation shows that $\sigma$ is precisely the coefficient that couples to the interface length, $\ell_\gamma$, and hence is a proper interfacial tension. Furthermore, since the two-dimensional lateral pressures are defined as
\be
p_\alpha = \frac{\partial F}{\partial A_\alpha} \,, 
\ee
we see that the pressure difference $\Delta p = p_+-p_-$ does indeed depend on the interfacial curvature. Although small - it is an $O(\xi)$ correction - this contribution is always present in phase coexistence of closed systems. It was first derived by Kelvin \cite{Thomson1872} from the Young-Laplace equation. 

\subsection{Coupling mechanisms between curvature and order parameter}
\label{sec:coupling}

Here we consider the most generic scenario in which all terms in Eqs. \eqref{F expansion} and \eqref{G variation}, including $k$ and $\bar{k}$, are non-vanishing. In this case the local curvature affects directly the magnitude of the order parameter $\phi$, instead of just indirectly influencing lateral displacement through non-trivial topology and intrinsic geometry. 

Without specifying the shape of $\Sigma$ nor the functional form of $k(\phi)$ and $\bar{k}(\phi)$ it is hard to make precise predictions. We will deal with a specific model and specific geometries in the next Section. Here we instead consider an approximately flat membrane, so we can treat the curvature terms as perturbations. If $k(\phi) H^2$ and $\bar{k}(\phi) K$ are much smaller than $f$, we get that the binodal points of the free energy are shifted by a small, curvature-dependent, amount. More precisely (see also Appendix \ref{app:linear}), we have that the Maxwell values are shifted as $\phi_\pm \to \phi_\pm + \delta \phi_\pm$, with
\be 
\delta \phi_\pm = \frac{\Delta k H^2 + \Delta \bar{k} K}{\phi_+ - \phi_- } - \frac{k(\phi_\pm) H^2 + \bar{k}(\phi_\pm) K}{f''(\phi_\pm)} \,,
\label{delta phi HK}
\ee
where $\Delta k = k(\phi_+) - k(\phi_-)$ and $\Delta \bar{k} = \bar{k}(\phi_+) - \bar{k}(\phi_-)$ are the differences between the bending moduli evaluated on the homogeneous binodal concentrations. Eq. \eqref{delta phi HK} shows how the equilibrium bulk phases are directly influenced by local curvature.

Since $\xi$ is smaller than any other scale, we can still assume that the interface separating the two phases lies entirely in a region where curvature can be considered to be constant along the $z$ geodesic normal direction. This implies that we can use again Eq. \eqref{line tension} to compute the line tension using the shifted binodal values \eqref{delta phi HK}, finding a curvature-dependent line-tension $\tilde{\sigma}$
\be
\tilde\sigma \simeq \sigma +  \delta_k H^2 + \delta_{\bar{k}} K + \dots 
\label{sigma HK}
\ee
where the dots stand for higher order terms in the curvatures. The two coefficients $\delta_{k,\bar{k}}$ are defined as integrals over the homogeneous miscibility gap
\be
\delta_{k,\bar{k}} = \xi \int_{\phi-}^{\phi_+}\D \varphi\,\frac{g_{k,\bar{k}}(\varphi)}{\sqrt{2 g(\varphi)}} \,,
\label{delta kkb}
\ee
where $g$ is defined in \eqref{shifted g potential} and $g_{k,\bar{k}}$ are defined in a similar manner, i.e. $g_{k,\bar{k}}(\phi_\pm)=g_{k,\bar{k}}'(\phi_\pm)=0$ and $g_{k,\bar{k}}''(\varphi)$ coincides with the second derivative of the bending moduli (see the derivation of equation \eqref{Zh 2} for more details). Interestingly, the terms $\delta_k/\sigma$ and $\delta_{\bar{k}}/\sigma$ in Eq. \eqref{sigma HK} resemble one-dimensional analogues of the Tolman lengths (see Sec. \ref{sec:effective_intro} and Ref. \cite{Tolman1949}).

If instead the curvature couplings are so small that they enter in the effective free energy $\mathcal{F}$ as $O(\xi)$ terms, they have a different effect. Formally, this can be achieved by  replacing $k(\phi) H^2+\bar{k}(\phi) K$ with $\xi (k(\phi) H^2+\bar{k}(\phi) K)$ in Eq. \eqref{F expansion} and Eq. \eqref{G variation}. This means that, contrary to the case we just discussed, the curvature interactions will not affect the interface profile Eq. \eqref{equipartition}, nor they will influence the line tension or the bulk phase values $\phi_\pm$. Rather, they will only affect equilibrium at $O(\xi)$, thus they will contribute to the determination of interface position. It is easy to show that in this case is equation \eqref{CGC} that needs to be modified to
\be
\sigma \kappa_g -\Delta k H^2  - \Delta \bar{k} K  = \mathrm{const}\,.
\label{JL interface}
\ee
Not surprisingly, this equation is precisely the one obtained by the first functional variation of the J\"ulicher-Lipowsky sharp interface model \cite{Julicher1993}, which we treated in detail in \cite{fonda2018interface}. 

This latter result hints at a more general concept. When adding environmental couplings to sharp interface models there is an implicit assumption about the subleading character of the interactions - relatively to an expansion in the interface thickness -, since they can affect the position of the interface but not its inner structure. Physical interfaces have however finite thickness, and thus any coupling with other degrees of freedom will naturally influence the interface as a diffuse thermodynamic entity, rather than just as a geometric submanifold. For this reason thin interface models, where $\xi$ is small but non-zero, can produce more physically reliable results.

\section{A simple model}
\label{sec:model}

\begin{figure*}
\includegraphics[width=\linewidth]{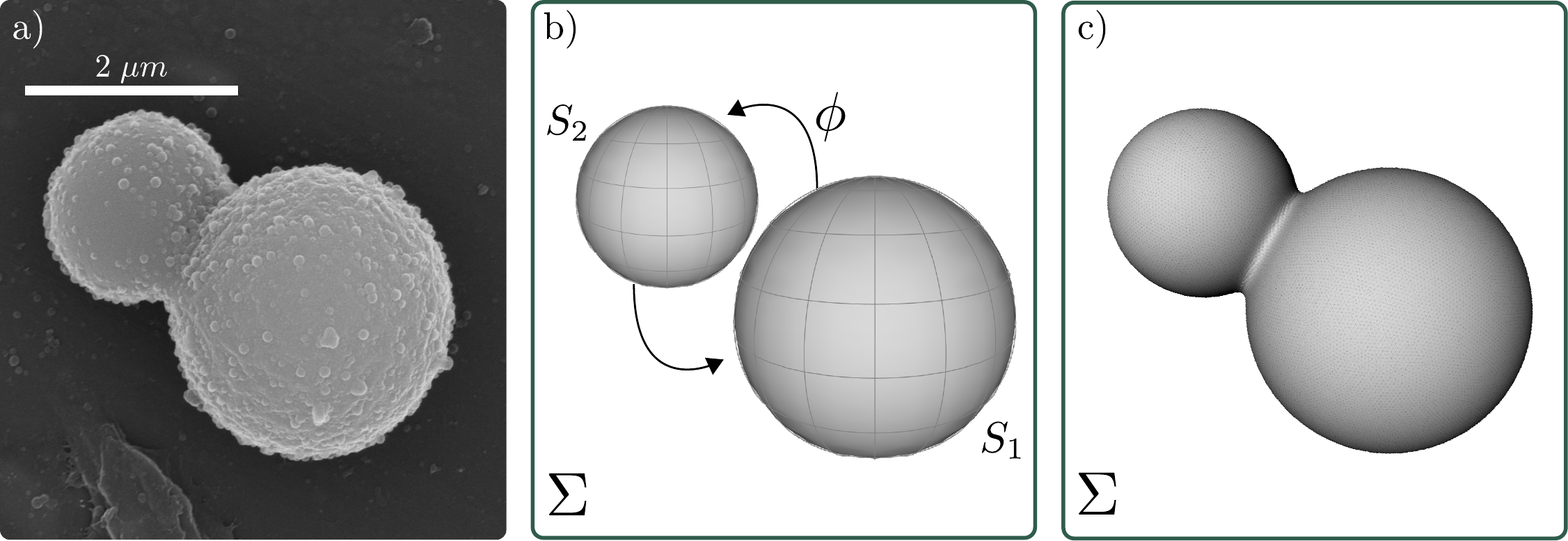}
\caption{\textbf{a)} Experimental scanning electron microscopy image of an asymmetric dumbbell-shaped colloid used as scaffold for the lipid membrane in SLVs (taken from \cite{rinaldin2018geometric}). \textbf{b)} To maintain analytical control over computations, we approximate the surface in a) as two disconnected spheres exchanging order parameter $\phi$ between each other but isolated from the environment. \textbf{c)} Full three-dimensional reconstruction of the dumbbell. The surface is an axisymmetric approximation of the image in a), with two spherical caps attached to a neck-like region obtained with a polynomial interpolation of order eight. It has area $\sim 17.11 R_1^2$, volume $\sim 5.40 R_1^3$ and Willmore energy $\sim 37.96$. The triangulated surface consists of 33131 vertices.}
\label{fig:spheres}
\end{figure*}

The rich phenomenology of binary mixtures on curved surfaces has much more to offer than the general results outlined in Sec. \ref{sec:effective}. To draw more precise conclusions, however, it is indispensable to make the $\phi-$dependence of the functions $D$, $f$, $k$ and $\bar{k}$ in Eq. \eqref{F expansion} explicit, and thus focus our analysis on a specific subset of possible material properties. Whereas this operation can be performed in multiple ways (see Sec. \ref{sec:effective_intro}), here we propose a simple and yet insightful strategy based on a curved-space generalization of the most classic microscopic model of phase separation, namely the lattice-gas model. 

To this purpose, we discretize $\Sigma$ into a regular lattice, with coordination number $q$ and lattice spacing $a$, this being defined as the geodesic distance between neighbouring sites. We ignore the fact that, for closed surfaces with genus $g\neq 1$, there are topological obstructions to construct regular lattices and point defects (i.e. isolated sites where the coordination number differs from $q$) become inevitable. We assume that these isolated points give a negligible contribution to the free energy in the continuum limit. Each site is characterized by a binary spin  $s_{i}=\pm 1$, serving as a label for either one of the molecular components (e.g. $s_{i}=+1$ indicates that the $i-$th site is occupied by a molecule of type $A$, while $s_{i}=-1$ in case the molecule is of type $B$). Because of the short range interactions between the molecules, the total energy of the system is computed via the Ising Hamiltonian:
\be
\mathcal{H} = - \sum_{\langle ij \rangle} J_{ij} s_{i} s_{j}  - \sum_{i} h_i s_i \;,
\label{hamiltonian}
\ee
where $i=1,\,2\ldots\,N$ and $\langle ij \rangle$ indicates a sum over all the pairs of nearest neighbours in the lattice. Finally, conservation of the total number of molecules implies:
\be
\sum_i  \left(\frac{1+s_{i}}{2}\right) = \Phi N\;.
\ee

Now, in the classic lattice-gas model, the coupling constant $J_{ij}$ and the external field $h_{i}$ are uniform across the system. Here, we allow them to depend on the local geometry of $\Sigma$. Using the same assumptions underlying the expansion \eqref{F expansion}, augmented by the additional symmetry $J_{ij}=J_{ji}$, yields:
\begin{subequations}
\label{ising_parameters}
\begin{align}
J_{ij} &= \frac{1}{4}\,\left(J + \qk \,\frac{H_{i}^2+H_{j}^2}{2} + \qkb\,\frac{K_{i}+K_{j}}{2}\right) \;, \label{ising quadratic} \\
h_{i} &= -\frac{1}{2}\,\left(\lk H_{i}^2 + \lkb K_{i}\right) \;,
\end{align}
\end{subequations}
where $H_{i}$ and $K_{i}$ are respectively the mean and the Gaussian curvature evaluated at the $i-$th lattice site. The $Q-$couplings modulate the relative strength of the attraction/repulsion between molecules, reflecting that both the distance and relative orientation of neighbouring molecules vary across the surface. Similarly, the $L-$couplings measure the propensity of a molecule to adapt to the local curvature. In particular, we note that $\lkb$ is exactly the only curvature coupling employed in Ref. \cite{paillusson2016phase} to describe the interaction of binary mixtures with minimal surfaces. We stress that, in order for the Hamiltonian \eqref{hamiltonian} to admit phase separation, $J_{ij}>0$. As the local Gaussian curvature can be both positive and negative, this is not necessarily true for a generic surface and an arbitrary choice of the constants $J$, $\qk$ and $\qkb$. In the following, we assume that $J>0$ is sufficiently large to prevent $J_{ij}$ from changing sign. Furthermore, we assume for simplicity all the other constants in Eqs. \eqref{ising_parameters} to be positive. The latter assumption is not indispensable and has not qualitative effects on the structure of the free-energy landscape and on the phase diagram.

The free energy of the mixture can now be easily calculated using the mean-field approximation, upon assuming the variables $s_{i}$ to be spatially uncorrelated (i.e. $\langle s_{i}s_{j} \rangle = \langle s_{i} \rangle \langle s_{j} \rangle$, with $\langle\cdots\rangle$ the ensemble average). Thus, letting
\begin{equation}
P(s_{i})=\phi_{i}\delta_{s_{i},1} + (1-\phi_{i})\delta_{s_{i},-1}\;,
\end{equation}
the probability associated with finding a molecule of type $A$ or type $B$ at $i-$th site, yields, after standard algebraic manipulations (see e.g. Ref. \cite{parisi1988statistical}),
\begin{multline}\label{eq:mean_field}
F = -\sum_{\langle ij \rangle} J_{ij}(2\phi_{i}-1)(2\phi_{j}-1)+\sum_{i}h_{i}(2\phi_{i}-1)\\	
+ T \sum_{i}\left[\phi_{i}\log \phi_{i}+(1-\phi_{i})\log(1-\phi_{i})\right]\;,
\end{multline}
with $T$ the temperature in units of $k_{B}$. Coarse-graining Eq. \eqref{eq:mean_field} over the length scale $\xi$, finally yields Eqs. \eqref{effective F} and \eqref{F expansion}, with 
\begin{subequations}\label{eq:lattice_gas}
\begin{align}
\label{ising D}
D(\phi) &= \xi^2 J \,, \\
\label{ising f}
f(\phi)  &=  T  \mathcal{S}(\phi) + q J \phi (1-\phi) \,,\\
\label{ising k}
k(\phi) &= q \qk \phi (1-\phi) + \lk \phi \,, \\
\label{ising kb}
\bar{k}(\phi) &= q \qkb \phi (1-\phi) + \lkb \phi \,,
\end{align}
\label{ising MF}%
\end{subequations}
where $\mathcal{S}(\phi) = \phi \ln \phi + (1-\phi)\ln (1-\phi)$ is the mixing entropy and we dropped $\phi$-independent terms from the bending moduli. The symmetry $\phi \leftrightarrow 1-\phi$ is explicitly broken only by linear $L-$couplings. Note that because of the total constraint on $\Phi$ we can disregard homogeneous terms linear in $\phi$, but we are not allowed to do the same for linear terms which depend on local geometry. 
Consistently with the assumptions about the separation of scales outlined in Sec. \ref{sec:effective} (i.e. $\xi^{2}H^{2} \sim \xi^{2} K \approx 0$), we have dropped curvature-dependent terms in the expression of $D$.

\subsection{Surfaces of constant curvature}
\label{sec:constant_curvature}
With Eqs. \eqref{eq:lattice_gas} in hand, we are now ready to fully explore the phase diagram of binary mixtures on curved surfaces. As a starting point, we consider the case of surfaces with constant curvatures, such as the sphere or the cylinder. In this case, the coupling of the order parameter with the curvatures, embodied by the third and second term in Eq. \eqref{F expansion}, merely results in a renormalization of the critical temperature. In fact, if $T > T_c$, with
\be
T_c = \frac{q}{2} \left( J + \qk H^2 + \qkb K \right) \,,
\label{ising Tc}
\ee
the free energy density $f(\phi)+k(\phi) H^2 + \bar{k}(\phi)K$ is always convex, and thus the homogeneously mixed configuration, $\phi=\Phi$, is the only stable equilibrium. 
Evidently, the linear terms in Eqs. \eqref{eq:lattice_gas} do not affect the convexity of the free energy, thus do not contribute to the critical temperature. 

Despite the known limitations of mean-field theory in two dimensions - here further corroborated by the experimental evidence that lipid mixtures belong to the same universality class as the two-dimensional Ising model \cite{Veatch2008,Honerkamp-Smith2008,Honerkamp-Smith2009} - it is nonetheless instructive to see how the generic picture illustrated in Sec. \ref{sec:curvature} specializes for the choice of potentials given by Eqs. \eqref{eq:lattice_gas} when $T \lesssim  T_c$ (which is the case for the majority of experiments on lipid membranes at room temperature).

At the first order in the Ginzburg-Landau expansion, the binodal concentrations are
\be 
\phi_\pm \simeq \frac{1}{2} \left(1 \pm \sqrt{3\,\frac{T_c-T}{T}} \right) \,.
\ee
From these we can compute the shifted potential $g(\varphi)$ of Eq. \eqref{shifted g potential}
\be
g(\varphi) \simeq \frac{4T}{3}\,(\varphi-\phi_+)^2(\varphi-\phi_-)^2 \,,
\ee
which is, as expected, a symmetric double-well quartic polynomial potential with minima at the binodal points. From here we can explicitly solve the interface profile equation \eqref{equipartition}, finding the well-known hyperbolic tangent kink
\be
\varphi(w) \simeq \frac{\phi_++\phi_-}{2} + \frac{\phi_+-\phi_-}{2} \tanh \left( \sqrt{2\frac{T_c-T}{J}}\,w \right) \,,
\label{kink}
\ee
where the zero of the geodesic normal coordinate $w$ (see the inset of Fig. \ref{fig:figure 1}) has been chosen such that the integral of the difference  $|\varphi-\phi_-|$ for $w<0$ matches the integral of $|\varphi - \phi_+|$ for $w>0$ (this is the definition of the Gibbs sharp interface, see Eq. \eqref{Gibbs 1}). 

The interface width, defined as the length scale over which $\phi$ changes from $\phi_-$ to $\phi_+$, scales as $\sim \xi J^{1/2} (T-T_c)^{-1/2}$ and diverges for $T \to T_c$. On the other hand the line tension can be computed to be
\be 
\tilde{\sigma} \simeq \xi \frac{1}{T} \sqrt{\frac{2}{J}}\,(T_c-T)^{3/2} \,,
\label{ising sigma tilde}
\ee
which instead vanishes at the critical temperature.  
With these results we can compute explicitly the quantities discussed in Section \ref{sec:coupling} when the curvatures are small. In particular, the curvature-dependent line tension can be evaluated using Eq. \eqref{delta kkb} - or equivalently by substituting  Eq. \eqref{ising Tc} into Eq. \eqref{ising sigma tilde} and expanding for small curvatures, finding 
\be
\frac{\delta_{k,\bar{k}}}{\sigma}
\simeq
q Q_{k,\bar{k}} \frac{3}{4}\, (T_c- T)^{-1}  \,.
\label{ising delta k}
\ee
Since this ratio is diverging for $T \to T_c$, it implies that curvature-dependent effects to the line tension, in our mean-field model, become more relevant near the critical temperature.

Similarly, if the curvature couplings are $O(\xi)$ and thus do not influence the interface profile nor the homogeneous binodal points, then the bending moduli differences of the J\"ulicher-Lipowsky model - as defined in Eq. \eqref{JL interface} - are
\begin{subequations}
\begin{align}
\label{MF Delta k}
\Delta k \simeq \lk \sqrt{\frac{3}{T}} (T_c-T)^{1/2}\,, \\
\label{MF Delta kbar}
\Delta \bar{k} \simeq \lkb \sqrt{\frac{3}{T}} (T_c-T)^{1/2} \,,
\end{align}
\label{MF Delta}%
\end{subequations}
which vanish at the critical temperature and depend only on the $L-$couplings since only terms that break the symmetry $\phi \leftrightarrow 1-\phi$ can produce a bending moduli difference. 

More generally, since the linear couplings $L_{k,\bar{k}}$ give no contribution to the redefinition of the critical temperature, Eq. \eqref{ising Tc}, nor to the line tension, Eq. \eqref{ising sigma tilde}, it might appear that they play no role in shaping the equilibrium phase diagram of the binary mixture. One would expect that adding a linear interaction term has no effect on the global thermodynamic stability of the system. In the next Section we will show how this is not the case when inhomogeneous surfaces are considered.

\begin{figure}
\includegraphics[width=\linewidth]{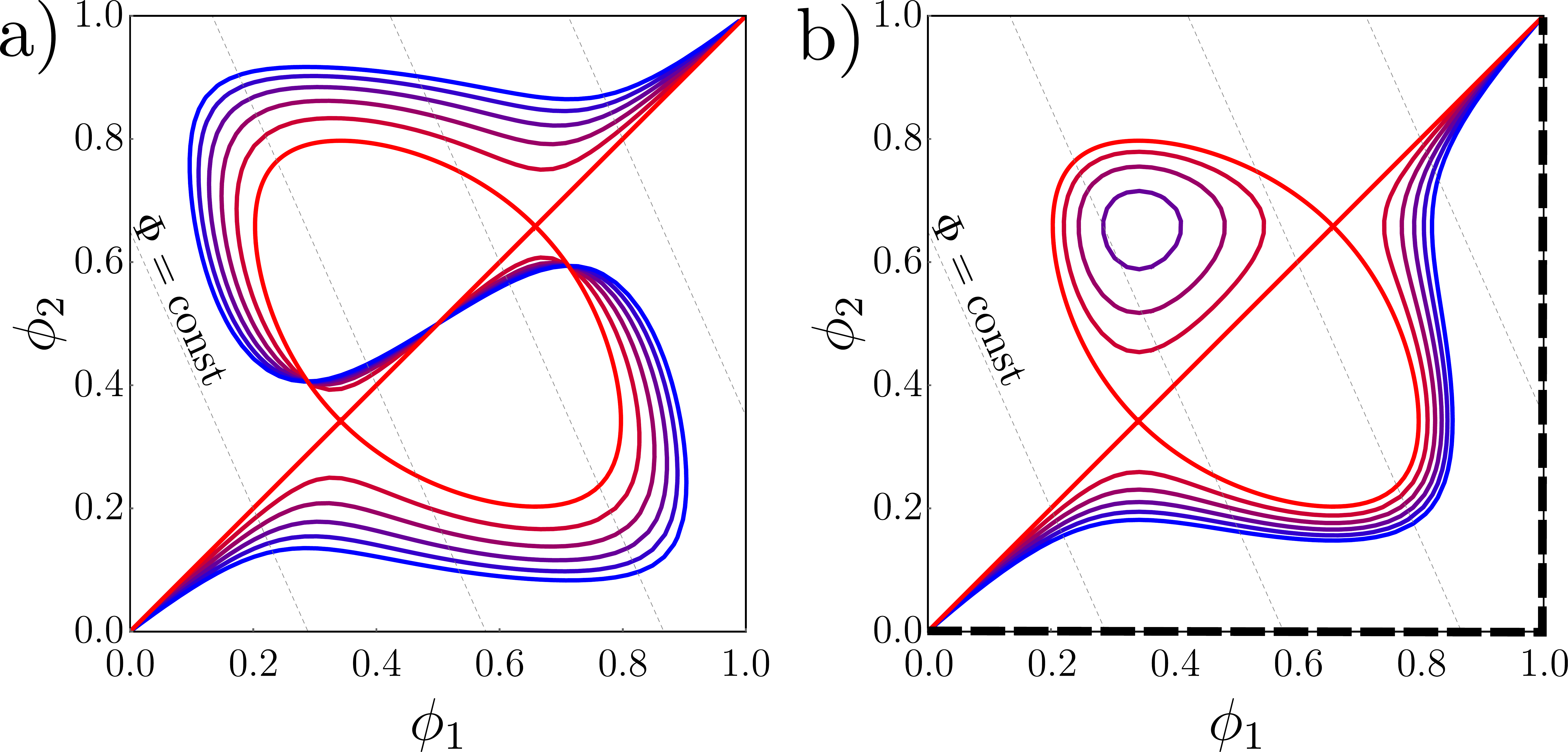}
\caption{
Lines of equilibrium.
Solutions of Eq. \eqref{two sphere eq} for two spheres with radii ratio $R_2/R_1=2/3$ at sub-critical temperature $T=0.45\,qJ$. \textbf{a)} lines  have all $T_L^{(a)}=0$ but regularly increasing $T_c^{(a)}$ from $1/2\,qJ$ (red) to $(1/2+1/10 R_a^{-2})\,qJ$ (blue). \textbf{b)} lines have all $\Delta T_c=0$ while $T_L^{(a)}$ increases from $0$ (red) to $1/20 \,R_a^{-2} qJ$ (blue). The black, thick, dashed line corresponds to the infinite $T_L^{(a)}$ limit. Note that the homogeneous solution $\phi_a=\Phi$ (the diagonal red line in both panels) is possible only in the absence of direct curvature couplings. Diagonal dashed lines are of constant $\Phi$. Notice that for a given $\Phi$ there can be multiple equilibrium solutions.
}
\label{fig:loeLM}
\end{figure}

\subsection{The phase diagram of disjoint spheres}
\label{sec:disjoint spheres}

In order to gain a deeper understanding of the role of curvature on the thermodynamics of phase separation, we need to consider a specific inhomogeneous shape. Building on our recent experimental results on SLVs \cite{rinaldin2018geometric}, we focus on asymmetric dumbbell-shaped substrates, as the one depicted in Fig. \ref{fig:spheres}a. 

In this case, $\Sigma$ consists approximately of two spherical caps connected to each other. We call the portion of the surface where the two spheres are in contact the ``neck region''. While the principal curvatures on the caps are approximately constant and proportional to their inverse radius, on the neck they reach higher values, so that both the mean and the (negative) Gaussian curvatures are significantly larger \footnote{In principle, the neck region could consist of a \textit{catenoid-}like shape, with $H \sim 0$. It is however sufficient a small deviation from this ideal case for $H^2$ to be much larger than $1/R_a^2$, the curvatures of the spherical caps.}. In terms of area, however, the neck occupies a relatively small portion of the whole surface. For the latter reason, in this Section we trade an accurate depiction of the geometry for analytic tractability and make the strong assumption that the neck will play a minor role in determining the equilibrium phase diagram of dumbbell-shaped two-dimensional liquid mixtures. Under this assumption, we approximate $\Sigma$ with a closed system consisting of two disjoint spheres, $S_1$ and $S_2$, of different radii, allowed to exchange molecules with one another, as shown in Fig. \ref{fig:spheres}b. Thus the total concentration can be expressed as
\be
\Phi  = \sum_{a=1,2} x_a \phi_a\,, 
\label{xi Phi}
\ee
where
$
\phi_a = 1/A_a \int_{S_a} \D A\,\phi\,, 
$
is the average concentration over the $S_{a}$ sphere ($a=1,\,2$), with $A_a = 4 \pi R_a^2$ the sphere area and $R_{a}$ the radius. Analogously, $x_a=A_a/A_\Sigma$, represents the area fraction of each sphere. Eq. \eqref{G variation} can now be solved using the mean-field parameter given by Eqs. \eqref{ising MF}, averaged over each sphere. Since for spheres $H^2=K=R^{-2}$, the four geometric couplings of Eqs. \eqref{ising MF} become pairwise equivalent, thus reducing the number of independent parameters to two: a symmetry-preserving quadratic term and a symmetry-breaking linear term. To see this explicitly we first minimize the free energy separately on each sphere, which gives the equations
\be 
1- 2\phi_a = \tanh \frac{T_L^{(a)}+2 T_c^{(a)} \left(1-2 \phi_a \right)-\mu}{2 T}\,,
\label{two sphere eq}
\ee
where we defined the local critical temperature by means of Eq. \eqref{ising Tc}
\be 
T_c^{(a)} = \frac{q}{2} \left(J + \frac{\qk + \qkb}{R_a^2} \right) \,,
\label{Tci}
\ee
and we introduced the curvature-dependent energy scale associated with the linear coupling
\be
T_L^{(a)} = \frac{\lk + \lkb}{R_a^2} \,.
\label{Tmi}
\ee
Constructing the equilibrium phase diagram of this system is a two-step process. First, one must find the values $\phi_a$ satisfying Eq. \eqref{two sphere eq} and the constraint \eqref{xi Phi}. Once these have been found, one must check the stability of each average concentration against spontaneous phase separation, i.e. verify whether $\phi_a$ lies within the local miscibility gap $[\phi_-^{(a)},\phi_+^{(a)}]$ on each sphere.

For fixed values of temperature and curvature couplings, the solutions of Eq. \eqref{two sphere eq} define a family of curves in the $\{\phi_1,\phi_2\}$ plane, as the total concentration $\Phi$ is smoothly changed from $0$ to $1$. We refer to these curves as ``lines of equilibrium'' and show some examples of them in Fig. \ref{fig:loeLM}. Although smooth, these lines do not need to be connected. Mathematically, they correspond to the set of points in concentration space where the gradient of the free energy is proportional to the vector $\{1,1\}$.

Fig. \ref{fig:loeLM}a shows the effect of varying the local critical temperature $T_c^{(a)}$ on each sphere. Since the free energy is still symmetric under the exchange $\phi \leftrightarrow 1-\phi$, the lines of equilibrium are invariant under the mapping $\phi_a \to 1-\phi_a$. Different colours correspond to different $\qk+\qkb$ values in Eq. \eqref{Tci}, ranging from $0$ (red) to $1/10\;qJR_1^{2} $ (blue). All curves pass through $\phi_1=\phi_2=1/2$.
Fig. \ref{fig:loeLM}b shows the effect of the linear $L-$couplings: $\lk+\lkb$ is increased from $0$ (red) to $1/20\;qJR_1^{2}$ (blue). In both panels the temperature is $T=9/2\;qJ$, and the spheres have radii $R_1=1$ and $R_2=2/3$.

It is instructive to compare, in closer detail, these results with those obtained in the absence of explicit coupling between the order parameter and the curvature, namely: $T_L^{(a)}=0$ and $T_c^{(1)}=T_c^{(2)}$ (the red-most lines in both panels). In this case, the lines of equilibrium consist of two mutually intersecting curves: a diagonal straight line $\phi_1=\phi_2=\Phi$, corresponding to the usual homogeneously mixed phase, and a second oval-shaped closed curve. The latter curve implies the existence of a second branch of solutions, where the amount of order parameter on each sphere is different from the total average. This result might be surprising, given that in this case the free energy density is homogeneous. However, it can be easily argued that this is an artefact of our model, originating from the following two arguments. First, the geometry we are considering is exceptional: the two spheres are not in direct contact and having $\phi_1 \neq \phi_2$ does not cost any extra interfacial energy, as it would be the case for a single connected surface. In fact, non-zero gradients would be strongly disfavoured. Secondly, it can be verified that the oval always lies within the miscibility gap of the potential and, therefore, even if mathematically possible, these extra solutions are  thermodynamically  metastable at best. This case alone shows another, rather general, fact: for a given set of external parameters, there can be multiple pairs of solutions of Eq. \eqref{two sphere eq}, each corresponding to a possible (meta-)stable equilibrium state. 

Spatial curvature changes this picture by introducing a smooth deformation of the lines of equilibrium. In Fig. \ref{fig:loeLM}a the straight line and the oval merge together into a single S-shaped connected curve, while in Fig. \ref{fig:loeLM}b one portion of the oval and of the straight line merge into a single line, and the rest splits into a closed curve. The latter becomes smaller and smaller as $T_L^{(a)}$ increases, and eventually disappears, leaving a single branch of equilibrium solutions. Our sign choices are such that it is thermodynamically preferable to first build-up non-zero $\phi$ on the largest sphere up to its maximum capacity (i.e. $\phi_{1}\approx\Phi$ and $\phi_{2}\approx 0$), rather than keeping the concentration everywhere uniform. Hence, at small $\Phi$, the lines of equilibrium bend towards the lower-right half of the diagram. For the linear coupling, this trend continues until the larger sphere is almost saturated. Then the concentration starts increasing on the small sphere too (so that the closed curves in the top left of Fig. \ref{fig:loeLM}b are always metastable). For the quadratic coupling the situation is more symmetric, in such a way that, for larger $\Phi$ values, it is more convenient to have a higher concentration on the small sphere. Note that, because of the classic double-well structure of the thermodynamics potentials, for a given $\Phi$ value there can be up to three different equilibrium solutions. Regardless of these quantitative differences, the main qualitative feature of the toy-model described in this Section is that, as a consequence of the influence of curvature on the free energy landscape of the binary mixture, the two disjoint spheres exhibit different concentrations despite being still in the ``mixed'' phase, i.e. without developing any interface. Interestingly, this phenomenon has some similarity with the thermodynamics of lipid membranes adhering onto non-homogeneous flat substrates \cite{Lipowsky2013}. 

\begin{figure*}
\includegraphics[width=\linewidth]{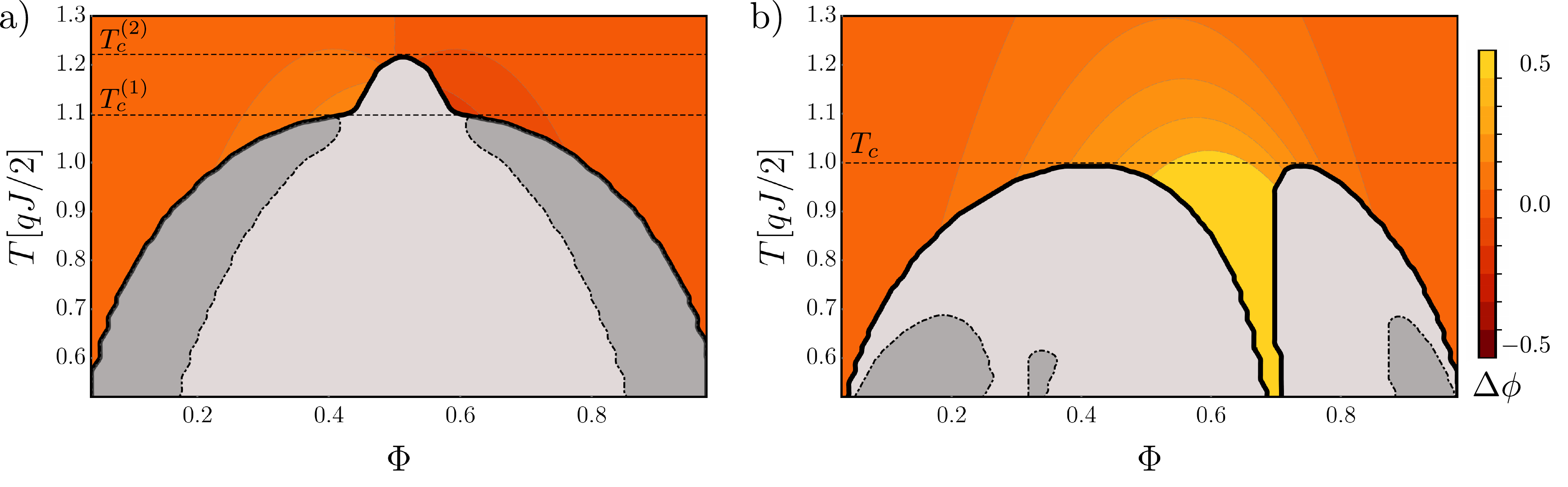}
\caption{Equilibrium phase diagrams in the presence of curvature interactions for the two-spheres system of Fig. \ref{fig:spheres}b with  $R_1=1$ and $R_2=2/3$. The solid black lines separate the generalized mixed phase (different shades of red/yellow, corresponding to degrees of inhomogeneity as shown in the legend) from the partially demixed phase (lighter gray). The dot-dashed line outlines the fully demixed phase (darker gray). All transition lines are binodals. \textbf{a)} Effect of the quadratic coupling, with $\qk+\qkb=1/5 \, R_1^2 qJ$. Each sphere has a different critical temperature, namely $T_c^{(1)}=0.55\, qJ$ and  $T_c^{(2)}=0.65\, qJ$. Since $\Delta T_L=0$, the diagram is symmetric for $\Phi \to 1-\Phi$. The inhomogeneity of the mixed phase is small and relevant only in the proximity of the critical temperatures. \textbf{b)} Effect of the linear coupling, with $\lk+\lkb=1/10\, R_1^2 qJ$. The critical temperature is the same for both spheres at $T_c=1/2 \,qJ$. The generalized mixed phase is strongly inhomogeneous in the region below $T_c$ and for concentrations $\Phi \sim x_1$.}
\label{fig:phase diagrams}
\end{figure*}

Fig. \ref{fig:phase diagrams} shows the phase diagram of the two-spheres system, obtained upon varying the temperature $T$ and the total concentration $\Phi$, while keeping $T_{c}^{(a)}$ and $T_L^{(a)}$ fixed. To highlight the specific role of each of these couplings, we isolate the effect of the quadratic coupling in Fig. \ref{fig:phase diagrams}a and that of the linear couplings in Fig. \ref{fig:phase diagrams}b, by setting $T_L^{(1)}=T_L^{(2)}$ and $T_{c}^{(1)}=T_{c}^{(2)}$ respectively. We see that there are essentially three stable phases (for the sake of simplicity, we focus only on stable phases and ignore metastable states): there is a mixed phase with no interfaces (red/yellow shades), there is a partially demixed phase with interfaces only on one sphere  (lighter gray), and finally there is fully demixed phase with phase separation occurring on both spheres (darker gray). To better characterize the mixed phase we introduce the difference
\be
\Delta \phi = \phi_1 - \phi_2 \,, 
\label{Delta phi}
\ee
which quantifies the departure of the concentration on a single sphere from the total average. A completely homogeneous mixed phase would then have $\Delta \phi=0$. The different shades of red/yellow in Fig. \ref{fig:phase diagrams} indicate different values of $\Delta \phi$, as shown in the legend. From the diagrams it is clear that, even in absence of genuine phase separation, one needs to relax and generalize the notion of mixing in order to grasp the complexity of the current scenario in comparison to the traditional picture. In fact, outside local miscibility gaps the ``mixed'' phase has a non-zero $\Delta \phi$. This effect is enhanced when there is a linear coupling, as in Fig. \ref{fig:phase diagrams}b, especially below $T_c$ and for concentrations close to the relative area ratio of the two spheres, $\Phi \sim x_1$ (which is equal to $\sim 0.69$ in the Figure).

Before dwelling into a detailed description of this phenomenon, let us emphasize that what we call here inhomogeneous mixing, is not a new thermodynamic phase, but rather the generalization of mixing to macroscopically non-homogeneous closed systems. In fact, the effect of inhomogeneities is smoothly smeared out at high temperatures, where the usual homogeneous mixing is always the true equilibrium. 

To see this, consider the limit where $T \gg T_c^{(a)}$ and $T \gg T_L^{(a)}$. We can then linearise the curvature couplings in Eq. \eqref{two sphere eq} and solve the equilibrium equation perturbatively. To this purpose, let us introduce the average critical temperature 
\be 
\hat{T}_c = \frac{T_c^{(1)}+T_c^{(2)}}{2} \,,
\label{hat Tc}
\ee
and the two energy scale differences
\be 
\Delta T_c = \frac{T_c^{(1)}-T_c^{(2)}}{2} \,,
\quad
\Delta T_L = \frac{T_L^{(1)}-T_L^{(2)}}{2} \,.
\label{Delta TcM}
\ee
By expanding Eq. \eqref{two sphere eq} at first order in $\Delta T_c$ and $\Delta T_L$, we get the deviation of the local concentrations from the total average  
\be
\Delta \phi = C_Q(\Phi,\hat{T}_c/T) \frac{\Delta T_c}{T} - C_L(\Phi,\hat{T}_c/T) \frac{\Delta T_L}{T} \,,
\label{small Dphi}
\ee
where $C_Q$ and $C_L$ are derived exactly in Appendix \ref{app:CLCM}. Their only relevant property is that they take finite values in the large $T$ limit, namely
\begin{subequations}
\begin{align}
\label{CL T infinity}
C_Q(\Phi,0) &= 4\Phi (1-\Phi) (2\Phi-1)\,, \\
\label{CM T infinity}
C_L(\Phi,0) &= 2	\Phi (1-\Phi)\,.
\end{align} 
\label{CLM T infinity}%
\end{subequations}
Eq. \eqref{small Dphi} clearly shows that, regardless of the magnitude of the curvature couplings, homogeneous mixing is always restored at high temperature. Furthermore, as the free energy is a continuous function of the concentrations $\phi_a$, such a crossover between inhomogeneous and homogeneous mixing occurs continuously, i.e. without passing through a first order phase transition. This argument can straightforwardly be extended to any arbitrary perturbative order in $\Delta T_{c}$ and $\Delta T_L$, demonstrating that equilibria with $\Delta \phi \neq 0$ and  $\Delta \phi = 0$ corresponds to different states of the same phase.

Despite of spatial curvature not giving rise to additional thermodynamic phases, its effect below the critical temperature is nonetheless dramatic as indicated by Fig. \ref{fig:phase diagrams}b. In this region of the phase diagram, the binodal line splits into two disconnected regions, separated by an intermediate continuum of states where $\Delta \phi$ is large and positive, hence the concentration on the two spheres is highly non-homogeneous. In a previous work, we have reported a direct experimental observation of these type of states and named the phenomenon ``antimixing'' \cite{rinaldin2018geometric}.

An intuitive understanding can be achieved by considering the limiting case in which $|\Delta T_L|$ overweights any other energy scale. Since the linear interaction breaks the $\phi \to 1-\phi$ symmetry, the energetic cost of having low or high concentrations of the order parameter become highly uneven and position-dependent. Specifically, if $\lk+\lkb$ is large and positive, with $R_1 > R_2$, having $\phi_2 \neq 0$ will cost much more energy than a non-zero concentration on $S_1$. Thus, any increment of the total concentration $\Phi$ will be first accommodated by $S_{1}$ until saturation (i.e. $\phi_{1}=1$) and only later the order parameter will start propagating on $S_{2}$. The corresponding lines of equilibrium associated with this scenario are represented as thick dashed black lines in Fig. \ref{fig:loeLM}b and consist of two perpendicular segments. The horizontal segment, i.e. $\{0\le \phi_{1} \le 1$,\,$\phi_{2}=0\}$, represents the build-up of order parameter on the sphere $S_{1}$, whereas the vertical segment, i.e. $\{\phi_{1}=1,\,0\le \phi_{2} \le 1\}$., indicates the subsequent build-up of order parameter on the sphere $S_{2}$.

\begin{figure}
\includegraphics[width=\linewidth]{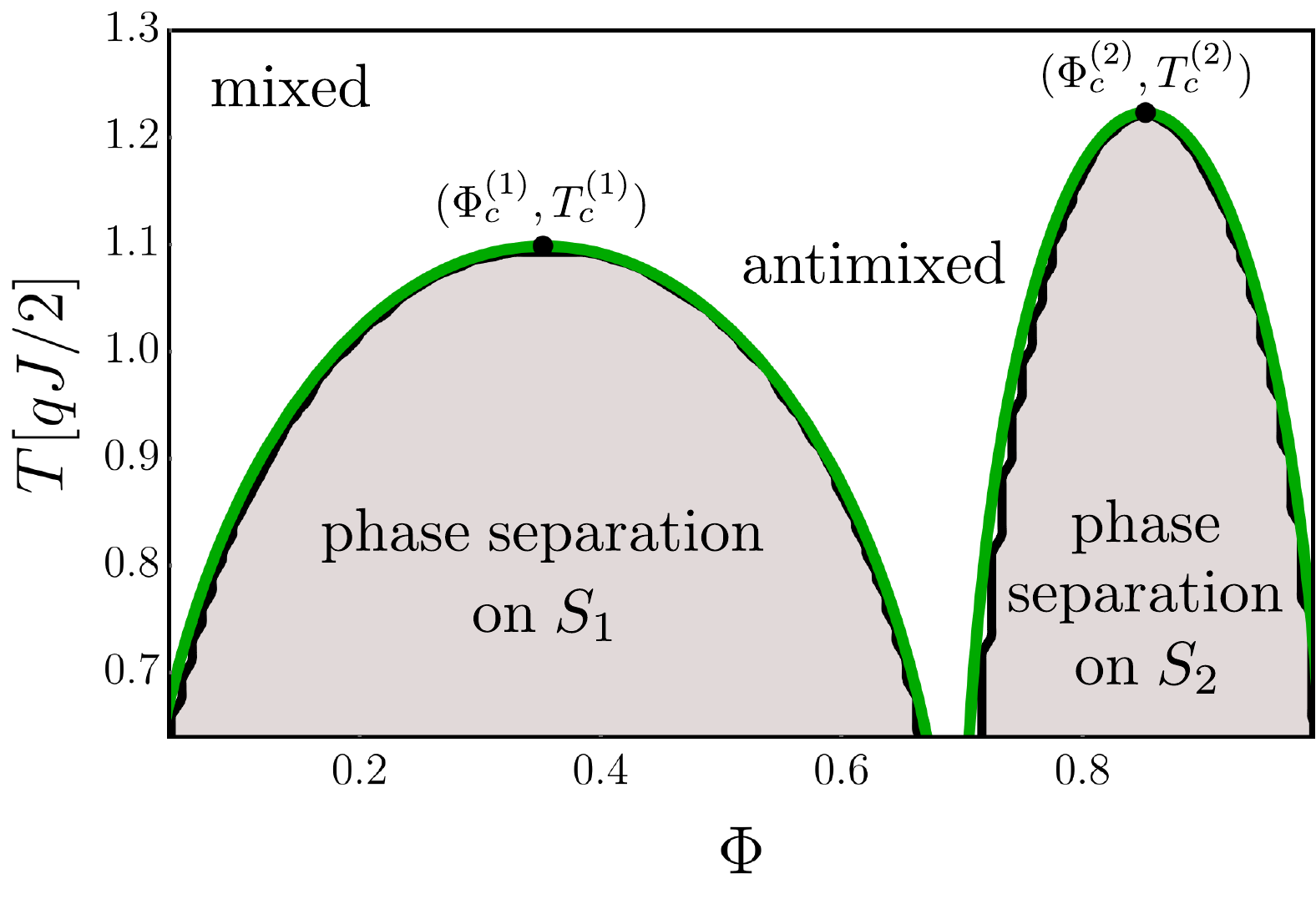}
\caption{The equilibrium phase diagram in the strong linear coupling limit. This figure is analogous to  Fig. \ref{fig:phase diagrams}, with non-zero linear and quadratic couplings: we set $\qk+\qkb=1/5 \, R_1^2 qJ$ and $\lk+\lkb=7/4 \, R_1^2 qJ$. The critical temperatures on each sphere are $T_c^{(1)}=0.55\, qJ$ and  $T_c^{(2)}=0.65\, qJ$. The green lines are the analytic binodal lines obtained from Eq. \eqref{T anitimixing binondal}. In the lighter gray region where phase separation happens on $S_1$, we have $\phi_2=0$. Conversely, in the region where phase separation happens on $S_2$, we have $\phi_1=1$. 
The two black dots are the critical points relative to each sphere, with critical concentrations given by Eq. \eqref{critical P antimixing}.}
\label{fig:antimixing}
\end{figure}
 
In this limit, the overall phase diagram is simply a disjoint union of the phase diagrams of each subsystem, given the simple mapping between $\phi_a$ and $\Phi$. This is illustrated in Fig. \ref{fig:antimixing}. Analyzing the stability of the mixed phase on each sphere is straightforward and leads to the conclusion that global phase separation is impossible in such a large $\Delta T_L$ limit, since there is no overlap between miscibility gaps of the two spheres. Moreover, the binodals of each sphere (the green lines in Fig. \ref{fig:antimixing}) can be analytically derived:  
\begin{subequations}
\begin{align}
T_{\rm binodal}^{(1)} &= \mathcal{T}\left(\frac{\Phi}{x_1}\right) \,,\\
T_{\rm binodal}^{(2)} &= \mathcal{T}\left( \frac{\Phi-x_1}{x_2}\right)  \,,
\end{align}
\label{T anitimixing binondal}%
\end{subequations}
with $\mathcal{T}(y) = (1-2y)/\mathrm{arctanh}(1-2y)$. Clearly, there are two distinct critical points of the system, specific for each sphere, located at $\{\Phi_c^{(a)},T_c^{(a)}\}$ in the phase diagram. The associated two critical temperatures are given by Eq. \eqref{Tci}, while the critical concentrations are
\be 
\Phi_c^{(1)} = \frac{x_1}{2} \;\;,\quad \Phi_c^{(2)} = x_1 + \frac{x_2}{2} \,.
\label{critical P antimixing}
\ee
With this analytical results it is then possible to give a precise definition of the antimixing phenomenon first reported in Ref. \cite{rinaldin2018geometric}: we define as {\em antimixed} the mixed phase of an inhomogeneous binary fluid at sub-critical temperature with non-overlapping local miscibility gaps.

Now, from a strictly technical point of view, it may be argued that our treatment of the substrate geometry is oversimplified, as we approximate the dumbbell-shaped membrane of Fig. \ref{fig:spheres}a with the two disjoint spheres of Fig. \ref{fig:spheres}b. Evidently, a real membrane is a single structure, and having $\Delta \phi \neq 0$ will inevitably induce gradients in the neck region that interpolates between the two lobes. Could these interfacial effects destroy the antimixed state? This question is addressed in the following Section.

\subsection{Numerical results on more general surfaces}
\label{sec:numerics}

In this Section we test whether our predictions on the existence of inhomogeneous mixing and antimixing hold for more realistic geometries. In particular, we must verify whether these bulk equilibrium states are compatible with the existence of concentration gradients. Therefore, let us now consider a new axisymmetric approximation of Fig. \ref{fig:spheres}a, i.e. the rotationally invariant surface of Fig. \ref{fig:spheres}c. Its radial profile has been obtained by joining two circular arcs by an interpolating polynomial of degree eight, chosen such that the neck interpolation and the circular arc match smoothly up to the fourth derivative at each of the two gluing points.

Our general strategy to find the equilibria is to implement an evolution equation that lets an arbitrary configuration smoothly flow towards minima of Eq. \eqref{grand canonical G}. Inspired by \cite{Rubinstein1992} we choose to implement gradient flow with conserved global order parameter:
\be
\partial_t \phi 
= - \frac{\delta G}{\delta \phi}
=  D \nabla^2 \phi - f'(\phi) - k'(\phi) H^2 - \bar{k}'(\phi) K + \mu\,,
\label{explicit flow}
\ee
where $\phi=\phi(\bm{r},t)$ is now a function of both space and flow parameter $t$. We stress that the $L^2$-gradient flow generated by Eq. \eqref{explicit flow} is purely fictitious and does not reflect the actual coarsening dynamics the binary fluid. However, this approach offers a practical way to generate stable equilibrium configurations for arbitrary geometries.

We then solve Eq. \eqref{explicit flow} numerically using a finite difference scheme on unstructured triangular meshes. More details about our numerical methods can be found in Ref. \cite{rinaldin2018geometric} and our code is available for download on GitHub \cite{githubMembrane}. Meshes are constructed using the software package \textit{Gmsh} \cite{geuzaine2009gmsh}. As in the case of planar droplets on the plane, the rotational symmetry of the substrate is not necessarily inherited by the minimizers of the Gibbs free energy $G$, thus it is necessary to solve the full two-dimensional problem.

\begin{figure}
\includegraphics[width=\linewidth]{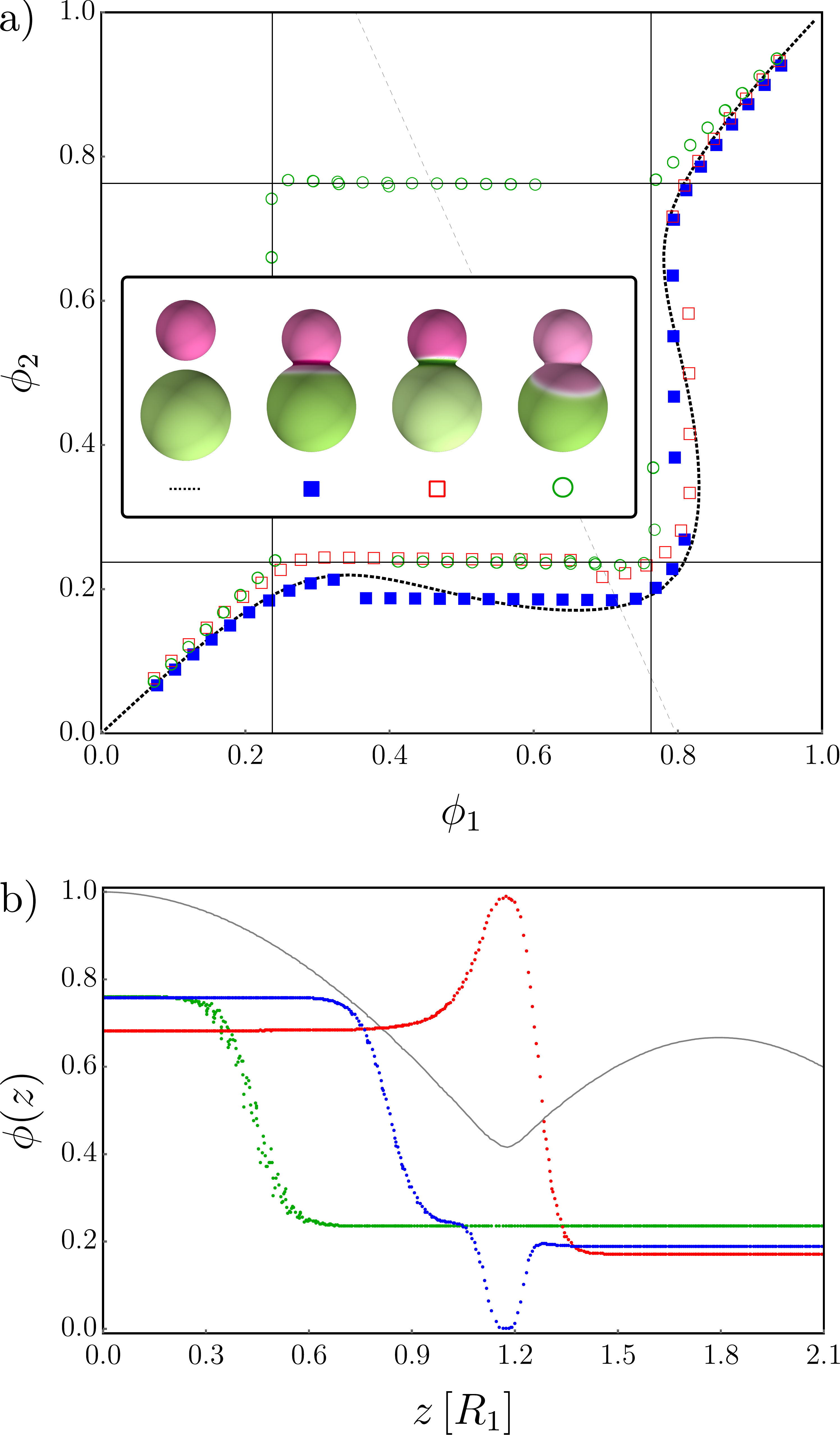}
\caption{
Equilibrium states of the axisymmetric geometry of Fig. \ref{fig:spheres}c, obtained from numerical solutions of Eq. \eqref{explicit flow} with mean-field potentials from \eqref{ising MF} at sub-critical temperature $T=.45\,qJ$. \textbf{a)} Lines of equilibrium for different $L-$couplings: the solid blue squares have $\lk=1/40\,qJR_1^2$, the empty red square have $\lkb=1/40\,qJR_1^2$ and the empty green circles have no direct interactions. The dashed black line is the line of equilibrium obtained as in Fig. \ref{fig:loeLM} for the two-spheres geometry with $\lk+\lkb=1/40\,qJR_1^2$. The solid vertical and horizontal lines are the binodal concentrations $\phi_\pm$ at zero coupling.	The inset shows what the four different equilibria look like at the same total concentration $\Phi=0.55$ (shown as a dashed gray line in the main plot). \textbf{b)} Concentration profiles as function of the arc-length axisymmetric coordinate $z$, for the three dumbbells shown in the inset of \textit{a)}. The thin black line in the background shows the radial profile of the surface (in cylindrical arc-length coordinates, the profile of a sphere looks like a trigonometric sine). The two horizontal dashed lines correspond to the Maxwell values $\phi_\pm$. In all simulations we set $\xi=0.024 \, R_1$.
}
\label{fig:comparison}
\end{figure}

Our main numerical results are shown in Fig. \ref{fig:comparison}a. We focus on the linear couplings that explicitly break the $\phi \rightarrow 1-\phi$ symmetry of the free energy, since they offer the most interesting phenomenology. 
In all the simulations summarized in Fig. \ref{fig:comparison}, we set the temperature to $T=0.9 T_c$ with $T_c = qJ/2$ uniform over all $\Sigma$. As a guide to the eye, the numerical data are superimposed to the stable branch of the lines of equilibrium associated with the two disconnected spheres (see also Fig. \ref{fig:loeLM}b), with $\lk+\lkb=1/40 \, q J R_1^2$. This value is almost an order of magnitude lower than the one used to construct the phase diagram of Fig. \ref{fig:phase diagrams}b, yet it can be shown that it retains antimixed states as equilibrium solutions. Each data point is obtained upon averaging the numerically found stationary solutions of Eq. \eqref{explicit flow} over ten random initial field configurations. To facilitate the comparison, the $\phi_a$ values are computed by integrating $\phi$ over axisymmetric regions which have the same area fraction $x_1$ as the one occupied by $S_1$ in the case of the two disjoint spheres. The solid horizontal (vertical) lines correspond to the Maxwell values $\phi_\pm$ on the small (large) sphere. In general, if the local concentrations take the binodal values, $\phi_a = \phi_\pm$, it means that the system is likely phase separated, with the interface lying in only one sub-region of $\Sigma$. 

The differently colours denote different values of $\lk$ and/or $\lkb$, while keeping the $Q-$couplings to zero. Different data points with the same colour correspond to different values of $\Phi$. The green circles corresponds to the homogeneous case, where  also $L_{k,\bar{k}}=0$, and demixing occurs uniformly over the entire surface. Outside of the binodal interval, i.e. for either $\Phi < \phi_-$ or $\Phi > \phi_+$, the equilibrium state is homogeneously mixed with $\phi_{1}=\phi_{2}=\Phi$ and the data points are aligned along the diagonal. Conversely, phase separation occurs when $\phi_{-}\le \Phi \le \phi_{+}$, for which the data points depart from the diagonal and either $\phi_1$ or $\phi_2$ - the one containing no interfaces - coincide with $\phi_\pm$.

The square dots correspond to either $\lk$ (full blue) or $\lkb$ (empty red) equal to $1/40 \,q J R_1^2$, with all other couplings set to zero. In both cases we find that the numerical results follow qualitatively the dashed line of equilibrium. The  coupling with the squared mean curvature, $\lk$, seems to be the one that follows the two disjoint sphere results more closely, and is the only one of the two data sets that features configurations with $\phi_1 > \phi_+$ and $\phi_2 < \phi_-$ (see the bottom-right corner of Fig. \ref{fig:comparison}a). 

Interestingly, for some $\Phi$, the equilibrium concentrations depart from a line of equilibrium and follow the horizontal (or vertical) binodal line, although only in a specific range of parameters (e.g. red dots, with $\Phi<0.5$) the data exhibit $\phi$ values that approximate the binodal value $\phi_{-}$ with reasonable accuracy. In all other cases, $\phi_{a}$ relaxes toward different $\Phi$-independent values. This behaviour likely originates from one or both of the following features of our model. First, the interpretation of $\phi_{a}$ is less stringent when applied to a connected dumbbell, where the two lobes are not geometrically distinct regions. Second, there might be additional contributions resulting from the finite thickness of the interface ($\xi=0.024R_1$ in Fig. \ref{fig:comparison}).
In general, these observation indicate that, in non-homogeneous spaces, the definition itself of phase-separation requires special care.

This latter statement can be made more precise by considering the inset of Fig. \ref{fig:comparison}a and Fig. \ref{fig:comparison}b. 
In both plots, $\Phi=0.55$, corresponding to the dashed diagonal line in Fig. \ref{fig:comparison}a.
This value lies within the miscibility gap, thus, in the absence of an explicit coupling with the curvature, the system phase separates, and since $\Phi \neq x_a$, the expected areas occupied by the two phases do not match the relative size of the two lobes, so that the interface will lie away from the dumbbell's neck. The snapshots in the inset of Fig. \ref{fig:comparison}a are color-coded based on the local $\phi$ value, with $\phi=0$ in magenta, $\phi=1$ in green and $\phi=1/2$ in white. The rightmost snapshot illustrates the case of homogeneous phase-separation with the associated interface lying along a constant geodesic curvature line, as predicted by Eq. \eqref{CGC} for homogeneous potentials. 

Fig. \ref{fig:comparison}b shows a plot of $\phi$ along a meridian as a function of the arc-length $z$ from the equator of the larger sphere. The green dots show the interfacial profile of the classical phase-separated configuration, i.e. the hyperbolic tangent kink, given by Eq. \eqref{kink}, interpolating between $\phi_+$ and $\phi_-$. The dots are not perfectly aligned since the interface itself is not axisymmetric, so the arc-length $z$ does not match exactly the geodesic normal coordinate we employed in Sec. \ref{sec:curvature}.  When either $\lk$ (blue dots) or $\lkb$ (red dots) are switched on, the configuration of the phase field $\phi$ changes dramatically. The field now interpolates between values which are {\em not} the binodal values - shown as two horizontal dashed lines in the plot - a signature of the fact that the curvature affects the bulk concentration, even away from high curvature regions. The influence of the curvature becomes particularly striking in the neck region, where the Gaussian and mean curvature couplings give rise to opposite effects. Since $H^2$ is always positive, the coupling $\lk \phi H^2>0$ favours small $\phi$ values in regions of high curvature, as demonstrated by the prevalence of magenta tones around the neck. Conversely, since $K<0$, the coupling  $\lkb \phi K$ is negative and favours higher $\phi$ values, as indicated by the prevalence of green tones. This behaviour is reversed in case $\lkb<0$, but, given the arbitrariness in the definition of the field $\phi$, this does not change the qualitative picture.

Notice that in both cases the interface is slightly shifted away from the neck, a phenomenon which is reminiscent of the behaviour of phase domains in axisymmetric sharp interface models of free-standing lipid membranes \cite{Baumgart2005}. Interestingly, both the blue and green profiles are not monotonic functions of $z$. Finally, note that all three cases interpolate between different values in the two bulk regions: this proves that, in general, the distinction between inhomogeneous mixing and demixing is not well-defined: we choose to interpret the blue curve as the realization of the antimixed state (since the profile interpolates between values which are outside the local miscibility gap) on a single, connected, smooth geometry.

\section{Discussion}

In this work we investigated the thermodynamic equilibrium of two-dimensional fluids confined on closed spatially curved substrates. Our model is primarily intended to describe self-organization in scaffolded lipid vesicles (SLVs) \cite{rinaldin2018geometric}, i.e. self assembled lipid bilayers supported by arbitrarily shaped colloidal particles. 
Our results, however, are also immediately applicable to any other mixture forced to lay on a curved surface, such as in the case of coating and adsorption phenomena at liquid interfaces. 

We considered a binary mixture that can be characterized by a single scalar order parameter $\phi$. The generalization of the phenomena discussed here to the case of $n$-nary fluids is relatively straightforward.
Crucially, we focused on closed thermodynamical systems, i.e. systems where there is no exchange of $\phi$ with the surrounding environment. This implies that the average total concentration, $\Phi$, is an externally fixed parameter. Equilibrium states are found from minimization of the Gibbs free energy $G=F-\hat{\mu}\Phi$, where the chemical potential $\hat{\mu}$ is here set by the constraint on the total concentration.

In Sec. \ref{sec:effective_intro} we constructed the most general form for $F$, using only symmetry and scaling arguments, and identified four $\phi-$dependent parameters that, together with the total concentration $\Phi$, determines the equilibrium state of the system. These are: the compressibility $D$, the homogeneous free energy density $f$ and the two bending moduli $k$ and $\bar{k}$. In Sec. \ref{sec:review} we reviewed the classical theory of phase separation for coexisting liquids. In Sec. \ref{sec:curvature} and Appendix \ref{app:thin} we reviewed the boundary layer analysis of the thin interface limit of two-dimensional phase-field models on curved surfaces, without direct curvature interactions. We derived the two-dimensional versions of the Young-Laplace and of the Kelvin equations. 

In Sec. \ref{sec:coupling} and Appendix \ref{app:linear} we considered the case where the bending moduli are small and yet non-vanishing. Depending on their scaling with respect to $D$, they produce very different effects. In case $k,\bar{k} \sim O(\sqrt{D})$, the Young-Laplace equation is changed to the equilibrium equation of the J\"ulicher-Lipowsky model \cite{Julicher1993}, which we studied in detail in Ref. \cite{fonda2018interface}. If, on the other hand, the bending moduli are of the same order of $f$, curvature effects become more dramatic and can result in local shifts of the binodal concentrations and a spatial dependence in the line tension $\sigma$. The deviation of $\sigma$ from its flat space value is parametrized by two length-scales, which are the one-dimensional analogues of the Tolman lengths \cite{Tolman1949} for three-dimensional droplets. 

Although very general, the results of Sec. \ref{sec:effective}, have limited predictive power, since the $\phi-$dependence of the phenomenological parameters is left unspecified. In order to overcome this limitation, in Sec. \ref{sec:model}, we derived these parameters from the mean-field approximation of a microscopic lattice-gas model with curvature dependent interactions [see Eqs. \eqref{ising MF}]. We found that the curvature of the substrate directly affects the structure of the free energy landscape via four non-equivalent couplings, which either break or preserve the symmetry of the free energy under exchange of the two phases (i.e. $\phi\to1-\phi$). We refer to them respectively as $Q-$ and $L-$interactions.

Motivated by the experimental results we reported in Ref. \cite{rinaldin2018geometric}, we applied our model to dumbbell-shaped membranes, as shown in Fig. \ref{fig:spheres}. For simplicity, we first approximated this surface as consisting of two disjointed spheres, allowed to exchange order parameter, but otherwise isolated from the environment (see Sec. \ref{sec:disjoint spheres}). We found that $L-$interactions, which linearly couple with the order parameter $\phi$, favour inhomogeneous mixing, i.e. a single phase with non-uniform concentration across the system. For our simple two-sphere geometry, this implies that each sphere is characterized by a distinct $\phi$ value, depending upon the strength of the $Q-$ and $L-$couplings and the local curvature radius.

Exceptionally, for certain specific $\Phi$ values, such an inhomogeneously mixed phase remains stable even {\em below} the critical temperature. In this regime, the inhomogeneity becomes more severe and the two spheres exhibit a stark concentration difference, even though phase separation has not occurred. We named this peculiar phenomenon, that was observed in Ref. \cite{rinaldin2018geometric} experimenting with scaffolded lipid vesicles (SLVs), {\em antimixing}, to stress that, albeit still in the mixed phase, the equilibrium concentrations split on the two opposite sides of a local miscibility gap. Surprisingly, this behaviour depends on the linear couplings between the concentration and the local curvature (i.e. the $L-$coupling, in our notation), despite these not altering the Maxwell construction and being thermodynamically irrelevant in binary membranes confined on homogeneous substrates. This originates from the fact that, in the presence of sufficiently large geometrical inhomogeneities and sufficiently strong symmetry-breaking coupling with the curvature, the phase diagram partitions into two sub-diagrams, each with its own distinct critical point (see Fig. \ref{fig:antimixing}).

Lastly, in Sec. \ref{sec:numerics}, we verified that inhomogeneous mixing and antimixing persist also on more realistic dumbbell-shaped substrates, obtained by connecting two spherical caps with a smooth neck (see Fig. \ref{fig:spheres}c). In this case, inhomogeneous mixing demands the occurrence of sharp concentration gradients, whose structure is substantially different than that of standard interfacial profiles. Most importantly, the average concentrations on the spherical lobes, i.e. the regions away from the neck, differ from the binodal values, even if the thermodynamic potential has the same Maxwell concentrations everywhere. This phenomenon is somewhat similar to the change in the bulk lateral pressure due to curved interfaces, as predicted by the Kelvin equation, whereas now the ambient curvature is inducing this change. Finally, we found that the $\lk$, i.e. the linear interaction with the squared mean curvature, produces equilibrium concentrations which match very closely the values found from the two-spheres simplified geometry, thus indirectly confirming that the antimixed state is a valid concept also for connected geometries.
  
\subsection*{Beyond mixing and demixing}

We have demonstrated that the thermodynamics of mixtures confined on inhomogeneous {\em closed} substrates, entails a spectrum of interesting phenomena that is, perhaps, broader than initially thought. In particular, the importance of closeness (i.e. the fact that a mixture cannot exchange material with the external environment), might have been overlooked in the past, even though, after the seminal work by Baumgart {\em et al.} \cite{Baumgart2003}, the interplay between geometry and chemical composition in multicomponent membranes has become a subject of thorough theoretical and experimental investigations.

One of the most fundamental outcomes of our analysis is that curvature inhomogeneities force to relax the usual distinction between mixed and demixed phases, since now concentration gradients and interface-like structures can be induced by curvature rather than spinodal instabilities. The very existence of antimixing, on dumbbell-shaped substrates, provides a prominent example of stable equilibrium states which have features of both phases.

From a model-building perspective, this implies that extreme care must be used in choosing the functional form of the bending moduli profiles $k(\phi)$ and $\bar{k}(\phi)$, since, even the simplest interaction term (e.g. the linear coupling introduced by Markin \cite{Markin1981}), can produce highly non-trivial effects to the equilibrium phase diagram of closed systems. Furthermore, slightly different choices can lead to very different phenomenologies, thus negatively affecting the validity of a given model.

Some of our predictions appear amenable to a reasonably viable experimental verification. 
First, we have shown that $Q-$interactions may induce a curvature-dependent line tension and critical temperature. Even experiments on multicomponent spherical vesicles can potentially test this effect by searching for a possible dependence of $\sigma$ and $T_{c}$ on the vesicle's radius.
Furthermore, we recall that curvature terms were neglected in deriving $D$ from the lattice-gas model [see Eq. \eqref{ising D}] to comply with the general assumptions of Sec. \ref{sec:effective}. Lifting these assumptions yields in fact:
\be 
D= \xi^2 \left(J + \qk H^2 + \qkb K \right) \,,
\ee
which reveals a curvature dependence exactly analogous to $T_c$ in Eq. \eqref{ising Tc}, since the mean-field value of the nearest-neighbour interaction simultaneously affects both quantities. Note that this effect does not have any implications when only $L-$interactions are considered, thus our conclusions on the antimixed state are unchanged. However, this relation does predict that not only compressibility, but also the effective diffusion (e.g. as measured from photobleaching experiments) of lipids on a vesicle might depend on the vesicle size. To the best of our knowledge, neither one of these phenomena has yet been experimentally investigated.

\paragraph*{Acknowledgements} 
This work was supported by the Netherlands Organisation for Scientific Research (NWO/OCW), as part of the Frontiers of Nanoscience program (MR) and the VIDI scheme (LG,PF). 

\appendix 

\section{Geodesic normal coordinates}
\label{app:normal}

If the equilibrium configuration is in a phase-separated state, then $\phi$ develops linear interfaces. In this Appendix we show how to construct a set of coordinates which is adapted to the arbitrary shape of the system. We suppose that locally the effective free energy is homogeneous, so the two Maxwell values of the pure phases $\phi_\pm$ are well defined. We then define the interface $\gamma$ (the black curve in Fig. \ref{fig:figure 1}) as the level set 
\be 
\gamma = \left\{ \bm{r} \in \Sigma\; :\; \phi(\bm{r}) = \frac{\phi_++\phi_-}{2}\right\}\,.
\label{def interface}
\ee
The fact that we choose the average value between the two pure phase concentrations $\phi_\pm$ as defining the interface is purely conventional and does not carry any special meaning. Any other level set would work equally well.
Note that $\gamma$ in general will consist of multiple curves, which we assume to be not mutually intersecting. If $\partial\Sigma = \varnothing$, the curves will be closed. From now on we restrict to the case of the interface consisting of a single, closed and simple curve, although the generalization to multiple interfaces is straightforward.

We can parametrize $\gamma$ by its arc-length $s$, and define the tangent two-vector $T^i(s)\equiv\partial_s x^i(s)$ ($i=1,\,2$) and fix the normal $N^i(s)$ to consistently point in the $\Sigma_+$ domain. The geodesic curvature of the curve is defined as 
\be
\kappa_g =  T^i T^j \nabla_i N_j \,, 
\label{geodesic curvature}
\ee
with $\nabla_i$ the covariant derivative on $\Sigma$ (see  the Appendices of \cite{fonda2018interface} for much more detail on the theory of curves applied to linear interfaces). The arc-length condition implies that the norm of the tangent vector is constant when parallel-transported along $\gamma$, i.e. $T^i T^j \nabla_i T_j = 0$. The fact that also $N^i$ is of unit norm along $\gamma$ implies that $T^i N^j \nabla_i N_j = 0$. Orthogonality to $T^i$ implies $\kappa_g = - T^i N^j \nabla_i T_j$.

Note that $T^i$ and $N^i$ are two-vector fields which are defined only along the curve, so that we are allowed to take derivatives of them only along $T^i$, and not in directions normal to the curve. To this purpose, we need to extend the coordinate system away from $\gamma$. The most natural way to do so is to use geodesic normal coordinates. In a sufficiently small neighbourhood of $\gamma$ we associate to every point $P$ in $\Sigma$ the coordinate pair $(s,z)$, where $z$ represent the length of the (unique)
geodesic segment starting from $P$ and intersecting $\gamma$ orthogonally. The point where this intersection occurs defines the value of $s$ (see the inset of Fig. \ref{fig:figure 1}). To these coordinates we associate the two vector fields $t^i=\partial_s$ and $n^i=\partial_z$ with the defining properties
\be
t^i\vert_{z=0} = T^i \;,\quad n^i\vert_{z=0} = N^i,
\ee
and $n_i t^i = 0$ in the whole neighbourhood. Note however that $y^i$ is not unit normalized outside of $\gamma$: $s$ is the arc-length only of the $z=0$ line. On the other hand $n^i n_i=1$ throughout the whole patch. The induced metric on $\Sigma$ is diagonal
\be
h_{ij} = t_i t_j + n_i n_j \,.
\ee
With these definitions, the gradient of the scalar field $\phi(s,z)$ is
\be
\nabla_i \phi = t_i \phi_s + n_i \phi_z \,,
\ee
where $\phi_s = \partial_s \phi$ and $\phi_z = \partial_z \phi$. We are finally able to compute the action of the Laplace-Beltrami operator on a scalar function in the adapted frame
\be
\nabla^2 \phi  = t^i t_i \phi_{ss} + \phi_{zz} + \kappa \phi_{z} \,, 
\label{adapted LB}
\ee
with the vector norm computed with respect to the induced metric. 
There is no mixed term $\phi_{sz}$ because of the orthogonality of the coordinates and there is no $\phi_s$ term because of the geodesicity of $\partial_z$. Here $\kappa= t^i t^j \nabla_i n_j$ is the geodesic curvature of the $z=\rm{const}$ lines and satisfies 
\be
\kappa \vert_{z=0} = \kappa_g \,.
\ee
The $z$-dependence of $\kappa$ is non-trivial and for arbitrary geometries it cannot be computed explicitly. In the neighbourhood of the interface we can expand for small $z$ and use standard formulas for normal variations of geometric invariants (see e.g. Ref. \cite{Fonda:2016ine}), finding
\be 
\kappa = \kappa_g - z \left(\kappa_g^2 + K\right) + O(z^2) \,.
\label{kappa K}
\ee
In general higher order terms in $z$ become increasingly complicated, depending on derivatives of $K$ along $z$. However, in case $K=\rm const$ it is possible derive exact results. For instance, for a flat surface with $K=0$ it is easy to prove (see also e.g. Appendix A of Ref. \cite{Elder2001})
\be
\kappa = \frac{\kappa_g}{1+z \kappa_g} \,,
\ee
whereas for a sphere of radius $R$ (thus with $K=1/R^2$), we get
\be
\kappa = \frac{1}{R} \cot\left[\mathrm{arccot}(\kappa_g R) +\frac{z}{R} \right] \,.
\ee

\section{Thin interface limit}
\label{app:thin}

In this Appendix we review the technical details of the thin interface approximation $D = \xi^2 \ll A_\Sigma$, for equations of the form given by Eq. \eqref{equilibrium 2}. Since the diffusive length is small, we can look for perturbative solutions. Being the thickness of interface also $O(\xi)$, there are essentially two regimes to consider: the bulk phases where gradients are mild (the so-called outer region) and the interface itself where derivatives are unbounded (the inner region). At each order in $\xi$, the outer expansion provides the correct boundary conditions for the inner expansion.

We focus now on the inner expansion. Since the interface $\gamma$ is assumed to be a smooth curve, we can use the adapted coordinate system outlined in the previous Appendix. The normal coordinate will lie in an interval of the order $z \in [-\xi, \xi]$, with positive (negative) $z$ pointing along $\Sigma_+$ ($\Sigma_-$) domains. In this approximation, we expand the scalar field and the chemical potential as
\begin{subequations}
\begin{align}
\phi(s,z) &= \phi^{(0)}(z) + \xi \phi^{(1)}(z) + \dots \,,
\label{phi xi}\\
\hat{\mu} &= \mu^{(0)} + \xi \mu^{(1)} + \dots \,,
\label{mu xi}
\end{align}
\label{xi expansion}%
\end{subequations}
where the dots stand for $O(\xi^2)$ terms. 

As specified in the main text, we assume that over the interface the $z$-derivatives scale at most as $\xi^{-1}$.
The $O(1)$ inner expansion of the equilibrium condition is thus
\be 
f'(\phi^{(0)}) = \mu^{(0)} + \xi^2 \phi^{(0)}_{zz} \,.
\label{EL order 0}
\ee
Asymptotic matching with the bulk boundary conditions shows unsurprisingly that $\mu^{(0)}$ is precisely the chemical potential obtained by the common tangent construction, while $\phi^{(0)}$ approaches the bulk values $\phi_\pm$. We can rescale the geodesic normal distance by $\xi$ so that the variable $w=z/\xi$ spans approximately the full real line $w \in [-\infty,\infty]$. Defining $\varphi(w)\equiv \phi^{(0)}(w \xi)$ we can rewrite \eqref{EL order 0} as
\be
\varphi_{ww} = g'(\varphi) \,, 
\label{profile equation}
\ee
where $g$ is the physically equivalent, shifted potential 
\be 
g(\varphi) = f(\varphi) - \varphi \frac{f(\phi_+)-f(\phi_-)}{\phi_+-\phi_-} + \frac{\phi_- f(\phi_+)- \phi_+ f(\phi_-)}{\phi_+-\phi_-} \,,
\label{shifted g}
\ee
which satisfies the properties $g(\phi_\pm)=g'(\phi_\pm)=0$ and $g''(\varphi)=f''(\varphi)$. Equation \eqref{profile equation} can be multiplied by $\varphi_w$ and integrated - the choice of $g$ is such that the integration constant is zero - and one obtains the equipartition relation in the main text, namely Eq. \eqref{equipartition}, whose solutions are one-dimensional kinks. Without specifying $f$ it is not possible to solve further, however note that since $g$ and its first derivative approach zero in the limit $w \to \pm \infty$, we have that the decay towards $\phi_\pm$ of $\varphi$ is always exponential $|\varphi-\phi_\pm| \underset{w \to \pm \infty}{\sim} e^{\mp \lambda_\pm w}$ with decay lengths
\be 
\lambda_\pm = \frac{\xi}{\sqrt{f''(\phi_\pm)}} \,,
\label{lambdas}
\ee
which are diverging at critical points.

We now consider the next term in the inner expansion. Equation \eqref{equilibrium 2} at order $O(\xi)$, upon the substitution $z\to w/\xi$, reads
\be
f''(\varphi) \phi^{(1)} = \mu^{(1)} + \phi^{(1)}_{ww} + \kappa_g \varphi_{w} \,. 
\label{EL order 1}
\ee
We now multiply this equation by $\varphi_w$ and integrate over $w$. By using the identity
\begin{multline}
\int_{-\infty}^{+\infty}  \D w \left(\phi^{(1)}_{ww} - f''(\varphi) \phi^{(1)} \right) \varphi_w =\\
= \int_{-\infty}^{+\infty} \D w \left(\varphi_{www} - f''(\varphi)  \right) \phi^{(1)} = 0 \,,
\end{multline}
which follows from \eqref{profile equation} and $\varphi_w(\pm \infty)=\varphi_{ww}(\pm \infty)=0$, we find the relation
\be
\mu^{(1)} = \frac{\kappa_g Z}{\phi_+-\phi_-} \,,
\ee
which proves that equilibrium interfaces must be curves of constant geodesic curvature. In the above expression we defined 
\be
Z
=
\int_{-\infty}^{+\infty} \D w  (\varphi_w)^2 
=
\int_{\phi_-}^{\phi_+} \D \varphi  \sqrt{2 g(\varphi)} \,,	
\ee
where the last equality follows from \eqref{profile equation}.

By taking the limit $w \to \pm \infty$ of \eqref{EL order 1}, one finds the asymptotic relation for $\phi^{(1)}$
\be
\phi^{(1)}(\pm \infty) =  \frac{\mu^{(1)}}{f''(\phi_\pm)} =  \frac{\kappa_g Z}{(\phi_+-\phi_-)f''(\phi_\pm)} \,.
\ee
This result (which was also derived e.g. in \cite{Rubinstein1992}) is in striking contrast with the usual $O(\xi)$ matching condition for non-conserved order parameters: the chemical potential renders $\phi^{(1)}$ non-zero also in the bulk phases.

Having specified how to expand \eqref{equilibrium 2} perturbatively in powers of $\xi$ and having solved the equations \eqref{EL order 0} at $O(1)$ and \eqref{EL order 1} at $O(\xi)$, the last step is to evaluate the free energy on the equilibrium solutions. To this purpose we assume that, at finite $\xi$, $\Sigma$ is partitioned into three distinct regions: a strip $\gamma^{(\xi)}$ centered at $\gamma$ and of width $\sim 2\xi$ separating the two bulk domains $\Sigma_\pm^{(\xi)}$ which consist of $\Sigma_\pm$ with the half-strip region removed. The area of the strip is $\approx 2 \xi \ell_\gamma$ with $\ell_\gamma$ the length of the interface. The area of the two bulk domains is
\be 
\mathrm{Area}(\Sigma_\pm^{(\xi)}) = x_\pm A_\Sigma - \xi \ell_\gamma + O(\xi^2) \,.
\ee
Now integrals over the whole surface can be split into the sum of three terms: if $\mathcal{G}(\phi)$ is an arbitrary function of $\phi$ and its derivatives, its integral over $\Sigma$ can be computed as
\begin{multline}
\frac{1}{A_\Sigma} \int_\Sigma \D A \mathcal{G}(\phi) 
=
x_+ \mathcal{G}(\phi_+) + x_- \mathcal{G}(\phi_-)-
\\
-\frac{\xi \ell_\gamma}{A_\Sigma} \left(\mathcal{G}(\phi_+)+ \mathcal{G}(\phi_-)  - \lim_{\xi\to 0}\frac{1}{\xi} \int_{-\xi}^{+\xi} \D z \mathcal{G}(\phi^{(0)}) \right) + \\
+ \xi \mu^{(1)} \left( 
x_+ \frac{\mathcal{G}'(\phi_+)}{f''(\phi_+)} 
+ 
x_- \frac{\mathcal{G}'(\phi_-)}{f''(\phi_-)}
\right) + O(\xi^2) \,.
\label{average Oxi}
\end{multline}
There are two contributions of $O(\xi)$: one from the integration of $O(1)$ terms on the strip, the other from the $O(\xi)$ corrections to the bulk integrals. The integral in the second line can be evaluated by substituting $dz = \xi dw$ and integrating over the real line.

By picking $\mathcal{G}(\phi)=1$ one immediately recovers the general condition $x_++x_-=1$, which obviously does not take any correction. Instead, by picking $\mathcal{G}(\phi)=\phi$ one computes the total average concentration. In this case the second line of \eqref{average Oxi} vanishes, because of
\be 
\lim_{\xi\to 0}\frac{1}{\xi} \int_{-\xi}^{+\xi} \D z \phi^{(0)}(z) = 2 \phi^{(0)}(0) \,,
\label{Gibbs 1}
\ee
and of the definition of $\gamma$, \eqref{def interface}. This result contains however some degree of arbitrariness, since we defined the limit in \eqref{Gibbs 1} in a symmetric manner: any other choice of the location of the interface within the strip $\gamma^{(\xi)}$ would have led to a different value. This ambiguity is fixed in general by an appropriate shift of the zero-point of the geodesic normal coordinate in such a way that the following equality holds
\be
\int_0^{+\infty} \D w (\varphi-\phi_+) + \int_{-\infty}^0 \D w (\varphi - \phi_-) = 0\,,
\label{Gibbs 2}
\ee
which defines the so-called Gibbs interface. This condition states that the integrated  difference between inner and outer concentrations should match on both sides of the $z=0$ line. Formally, we should replace definition \eqref{def interface} with \eqref{Gibbs 2}, even if nothing of the following results depends on this choice.  We finally find that
\be 
\Phi = x_+ \phi_+ + x_- \phi_- +  \mu^{(1)} \left( x_+ \lambda_+ + x_- \lambda_- \right) \,,
\ee
i.e. the total concentration does indeed  pick a contribution from the interface and deviates from the homogeneous relation \eqref{total phi}. The extra factor depends on penetration depths defined in \eqref{lambdas}, and vanishes for geodesics.

By plugging $\mathcal{G}(\phi)=f'(\phi)$ into \eqref{average Oxi}, we precisely re-obtain the chemical potential expansion \eqref{mu xi}. Instead, by choosing $\mathcal{G}(\phi)=f(\phi)$ one finds
\begin{multline}
\int_\Sigma \D A f(\phi) =  \frac{1}{2} \sigma \ell_\gamma + \\ + A_\Sigma  \sum_{\alpha=\pm} x_\alpha \left( f(\phi_\alpha) + \sigma \kappa_g \frac{f'(\phi_\alpha)}{(\phi_+-\phi_-)f''(\phi_\alpha)} \right)\,,
\end{multline}
with $\sigma \equiv \xi Z$. Finally, with $\mathcal{G}(\phi)=\xi^2/2\nabla_i\phi \nabla^i \phi$ one finds
\be 
\frac{\xi^2}{2} \int_\Sigma \D A \nabla_i\phi \nabla^i \phi=  \frac{1}{2} \sigma \ell_\gamma  \,.
\ee
Combining the last two expression we obtain the $O(\xi)$ expansion for the total free energy, Eq. \eqref{thin F}.

\section{Linear corrections to the Maxwell construction}
\label{app:linear}

In this Appendix we show how to compute the linear corrections when the free energy $f(\phi)$ of Eq. \eqref{equilibrium 2} is modified by a small perturbation
\be
\tilde{f}(\phi) = f(\phi) + \varepsilon h(\phi) \,,
\ee
with $\varepsilon \ll 1$. In the following, we will keep only first order corrections in $\varepsilon$. The Maxwell common tangent condition reads
\be 
f'(\tilde{\phi}_\pm) + \varepsilon h'(\tilde{\phi}_\pm) = \frac{\tilde{f}(\tilde{\phi}_+)-\tilde{f}(\tilde{\phi}_-)}{\tilde{\phi}_+-\tilde{\phi}_-} \,,
\ee
where $\tilde{\phi}_\pm = \phi_\pm + \varepsilon \delta\phi_\pm + \dots$ are the shifted values of the bulk phases. The $O(\varepsilon)$ solution to these equations gives
\be 
\delta\phi_\pm = \frac{h(\phi_+)-h(\phi_-)}{\phi_+-\phi_-}- \frac{h(\phi_\pm)}{f''(\phi_\pm)} \,.
\ee
We now compute the $O(\varepsilon)$ correction to the line tension \eqref{line tension}. First, it is easy to see that the shifted potential $g$ defined in \eqref{shifted g} becomes 
\be
\tilde{g}(\phi) = g(\phi) + \varepsilon g_h(\phi)\,,
\ee
with
\be
g_h(\phi)=
h(\phi)
+
\frac{h(\phi_-)(\phi-\phi_+)-h(\phi_+)(\phi-\phi_-)}{\phi_+-\phi_-}\,,
\ee
which remarkably does not depend on $f(\phi)$ nor on its derivatives. Substituting the above expression in \eqref{line tension} and expanding again, one finds
\be
\tilde{\sigma} = \sigma + \varepsilon  \delta_h \,,
\label{eps sigma 1}
\ee
where
\be 
\delta_h = \xi \int_{\phi-}^{\phi_+}\D \varphi \frac{g_h(\varphi)}{\sqrt{2 g(\varphi)}} \,.
\label{Zh 1}
\ee 
Note that in \eqref{eps sigma 1} there are no endpoint contributions at order $O(\varepsilon)$ since $g(\phi_\pm)=0$. The integral in the above expression can be rewritten by means of the Gibbs condition \eqref{Gibbs 2}. Formally we can compute the integral of the linear and constant terms as
\begin{subequations}
\begin{align}
\int_{-\infty}^{+\infty} \D w\frac{h(\phi_-)(\phi_+-\varphi)}{\phi_+-\phi_-} &= \int_{-\infty}^0 \D w h(\phi_-)\,,
\label{eps Gibbs 1}\\
\int_{-\infty}^{+\infty} \D w\frac{h(\phi_+)(\varphi-\phi_-)}{\phi_+-\phi_-} &= \int_0^{+\infty} \D w h(\phi_+)  \,,
\label{eps Gibbs 2}
\end{align}
\label{eps Gibbs}%
\end{subequations}
\!where the integration limits should be thought as momentarily regularized. Plugging this into \eqref{Zh 1}, we find
\begin{multline}
\delta_h
=
\xi \int_0^{+\infty} \D w (h(\varphi)-h(\phi_+))
+\\+
\xi \int_{-\infty}^0 \D w (h(\varphi)-h(\phi_-)) \,,
\label{Zh 2}
\end{multline}
which shows how the first correction to the line tension is due to the integrated difference between the zero-th order inner and outer values of $h(\phi)$, evaluated on either side of the Gibbs interface. Any term which is symmetric with respect to the exchange $z\to -z$, such as constant and linear terms, will give a vanishing contribution to $\delta_h$. In the main text, we replace $\varepsilon$ by $H^2$ or $K$, and $h(\phi)$ by either $k(\phi)$ or $\bar{k}(\phi)$.

\section{High temperature expansion of the inhomogeneous mixing}
\label{app:CLCM}

Given the definition of $\Delta \phi$ in Eq. \eqref{Delta phi}, we can rewrite the local concentrations as
\begin{subequations}
\begin{align}
\label{Delta phi1 }
\phi_1 &= \Phi + x_2 \Delta \phi \,, \\
\label{Delta phi2}
\phi_2 &= \Phi - x_1 \Delta \phi\,,
\end{align}
\label{Delta phi12}%
\end{subequations}
and the local quadratic and linear couplings as 
\be
T_{c,M}^{(a)}= \hat{T}_{c,M} + (-1)^a \Delta \hat{T}_{c,M} \,,
\ee
where $\hat{T}_c$ is defined in Eq. \eqref{hat Tc}, $\hat{T}_L$ has an obvious analogous definition and $\Delta T_{c,M}$ are defined by Eqs. \eqref{Delta TcM}. By plugging these expressions into \eqref{two sphere eq} and expanding for small differences we get the equation
\be
\frac{T \Delta \phi}{\Phi (1-\Phi)} -  4 \Delta \phi \hat{T}_c + 4 (1-2\Phi) \Delta T_c  + 2 \Delta T_L = 0 \,,
\ee
whose solution is of the form \eqref{small Dphi} with coefficients
\be 
C_Q\left(\Phi, \hat{T}_c/T \right) = 4 \frac{\Phi (1-\Phi)(2\Phi-1)}{1-4 \frac{\hat{T}_c}{T} \Phi (1-\Phi)}\,,
\ee
and
\be 
C_L\left(\Phi, \hat{T}_c/T \right) = 2 \frac{\Phi (1-\Phi)}{1-4 \frac{\hat{T}_c}{T} \Phi (1-\Phi)}\,.
\ee
These are  always finite quantities whenever $T> 2 \hat{T}_c$, thus expansion \eqref{small Dphi} can be trusted only in the high temperature limit, where they approach the values of \eqref{CLM T infinity}.


%

\end{document}